%% file: sample-sigconf-authordraft.tex
\documentclass[sigconf]{acmart}
\AtBeginDocument{%
  }

\newcommand{\model}{RAID}
\usepackage{tikz}
\usepackage{amsmath}
\usepackage{mathrsfs}
\usepackage{comment}
\usepackage{algorithm}
\usepackage{algpseudocode}
\usepackage{booktabs} 
\usepackage{multirow} 
\usepackage{enumitem}
\DeclareMathOperator*{\argmin}{arg\,min}
\newtheorem{lemma}{Lemma}
\usepackage{xcolor}

\usepackage{filecontents}
\usepackage{subfigure}
\usepackage{pifont}

\setcopyright{acmlicensed}
\copyrightyear{2018}
\acmYear{2018}
\acmDOI{XXXXXXX.XXXXXXX}
\acmConference[Conference acronym 'XX]{Make sure to enter the correct
  conference title from your rights confirmation email}{June 03--05,
  2018}{Woodstock, NY}
\acmISBN{978-1-4503-XXXX-X/2018/06}

\begin{document}

\title{RAID: An In-Training Defense against Attribute Inference Attacks in Recommender Systems}

\author{Xiaohua Feng}
\email{fengxiaohua@zju.edu.cn}
\affiliation{%
  \institution{Zhejiang University}
  \city{Hangzhou}
  \country{China}
}

\author{Yuyuan Li}
\email{y2li@hdu.edu.cn}
\affiliation{%
  \institution{Hangzhou Dianzi University}
  \city{Hangzhou}
  \country{China}
}

\author{Fengyuan Yu}
\email{fengyuanyu@zju.edu.cn}
\affiliation{%
  \institution{Zhejiang University}
  \city{Hangzhou}
  \country{China}
}

\author{Ke Xiong}
\email{happywinder@zju.edu.cn}
\affiliation{%
  \institution{Zhejiang University}
  \city{Hangzhou}
  \country{China}
}

\author{Junjie Fang}
\email{junjiefang@hdu.edu.cn}
\affiliation{%
  \institution{Hangzhou Dianzi University}
  \city{Hangzhou}
  \country{China}
}

\author{Li Zhang}
\email{zhanglizl80@gmail.com}
\affiliation{%
  \institution{Zhejiang University}
  \city{Hangzhou}
  \country{China}
}

\author{Tianyu Du}
\email{zjradty@zju.edu.cn}
\affiliation{%
  \institution{Zhejiang University}
  \city{Hangzhou}
  \country{China}
}

\author{Chaochao Chen}
\authornotemark[1]
\email{zjuccc@zju.edu.cn}
\affiliation{%
  \institution{Zhejiang University}
  \city{Hangzhou}
  \country{China}
}

\renewcommand{\shortauthors}{Feng et al.}

\begin{abstract}
In various networks and mobile applications, users are highly susceptible to attribute inference attacks, with particularly prevalent occurrences in recommender systems. 
Attackers exploit partially exposed user profiles in recommendation models, such as user embeddings, to infer private attributes of target users, such as gender and political views. 
The goal of defenders is to mitigate the effectiveness of these attacks while maintaining recommendation performance. 
Most existing defense methods, such as differential privacy and attribute unlearning, focus on post-training settings, which limits their capability of utilizing training data to preserve recommendation performance. 
Although adversarial training extends defenses to in-training settings, it often struggles with convergence due to unstable training processes. 
In this paper, we propose \model{}, an in-training defense method against attribute inference attacks in recommender systems.
In addition to the recommendation objective, we define a defensive objective to ensure that the distribution of protected attributes becomes independent of class labels, making users indistinguishable from attribute inference attacks.
%
%
Specifically, this defensive objective aims to solve a constrained Wasserstein barycenter problem to identify the centroid distribution that makes the attribute indistinguishable while complying with recommendation performance constraints. 
To optimize our proposed objective, we use optimal transport to align users with the centroid distribution.
We conduct extensive experiments on four real-world datasets to evaluate \model{}.
The experimental results validate the effectiveness of \model{} and demonstrate its significant superiority over existing methods in multiple aspects.
\end{abstract}




\maketitle

\input{section/1-introduction}
\input{section/2-Background}

\input{section/3-Preliminary}
\input{section/4-Method}
\input{section/5-Evaluation}
\input{section/6-Discussion}
\input{section/7-Conclusion}





\appendix

\input{section/Appendix}

\end{document}

%% file: section/1-introduction.tex
\section{Introduction} \label{sec:intro}

Attribute Inference Attacks (AIA) pose emerging threats to user privacy across various application domains, ranging from recommender systems~\cite{otterbacher2010inferring,weinsberg2012blurme,wu2020joint}, to social media~\cite{zheleva2009join,chaabane2012you,gong2016you,gong2018attribute} and mobile platforms~\cite{michalevsky2015powerspy,narain2016inferring}. 
In such attacks, adversaries exploit exposed user profiles to infer private attributes, including location, gender, sexual orientation, and political views.
%
For instance, in recommender systems, the exposed profile could be a list of items rated by users, such as movies, mobile applications, and videos~\cite{weinsberg2012blurme,goga2013exploiting,shu2017user}. 
In social media, it might be the list of pages a user has liked on a social platform~\cite{kosinski2013private}. 
Attacks can utilize this profile to precisely infer sensitive attributes, leading to privacy leakages.
Notably, recommender systems rely heavily on 
sensitive interactions and contextual data to enhance the performance of recommendations. 
Particularly for new users (i.e., cold-start), recommender systems extensively integrate side information about users~\cite{adomavicius2010context,wang2020next,zhang2020gcn}. 
These factors further increase the risk of AIA for recommender systems in real-world scenarios.
Additionally, given that models in recommender systems nearly span those common models in social media and mobile platforms, defensive strategies against AIA developed for recommender systems can be easily extended to other domains. 
For example, incorporating Differential Privacy (DP) techniques into recommendation systems to protect users' privacy attributes can also be applied to social media platforms, thereby reducing the success rate of AIA. 
Based on the analysis above, we focus on exploring AIA within recommender systems in this paper.

Recommender systems, as effective solutions for combating information overload, have been widely adopted across various applications~\cite{rashid2002getting,konstan2012recommender,beigi2018similar}. 
These systems, through machine learning models, can automate the mining of user preferences and their relationships with interaction items. 
Due to its simplicity and effectiveness, Collaborative Filtering (CF)~\cite{koren2009collaborative,shi2014collaborative} is the predominant model used in recommender systems.
CF encodes user and item embeddings, thereby recommending similar items to users with similar behaviors.
%
%
%
%
Due to the 
correlation between user-item interactions and private user attributes~\cite{jia2018attriguard}, recommendation models may inadvertently encode sensitive information in their latent embeddings, even though the training data of recommendation models do not explicitly contain information about these private attributes.
Existing studies~\cite{weinsberg2012blurme,beigi2020privacy,guo2022efficient} show that if malicious attackers can access the parameters of the recommendation model or its recommendation results, they can accurately infer private user attributes.

\begin{figure}[t]
\centering
\includegraphics[width=0.95\columnwidth]{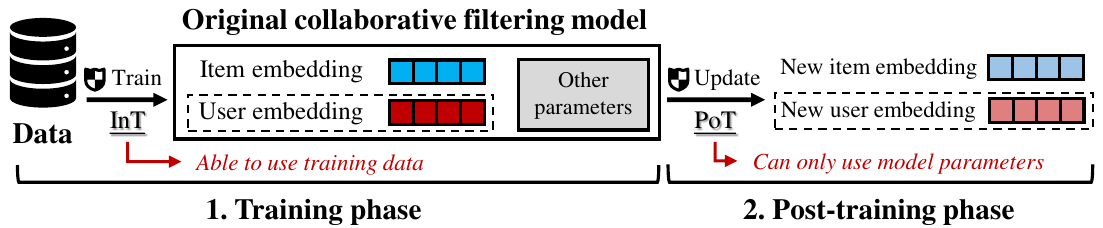} 
\caption{The difference between the settings of In-Training (InT) and Post-Training (PoT) primarily lies in their implementation stages and dependency on data. 
The InT setting is implemented during the model training phase, allowing it to protect user attributes while utilizing the original training data to preserve recommendation performance.
In contrast, the PoT setting is implemented after the model training is completed, at which point only the model's parameters are accessible~\cite{li2023making}.
The experimental results of the comparison are reported in Section~\ref{subsec:limit}.
}
\label{fig:diff}
\end{figure}

In response to AIA on recommender systems, existing studies propose various defense methods.
The primary goal of these methods is to effectively defend against AIA while ensuring that the recommendation performance remains unaffected. 
The main defense methods include techniques based on DP~\cite{mcsherry2009differentially,balu2016differentially,zhu2016differential}, attribute unlearning~\cite{guo2022efficient,li2023making}, and adversarial training~\cite{beigi2020privacy,ganhor2022unlearning}.
The DP methods employ a Post-Training (PoT) setting, introducing dynamic perturbations into the model's embeddings to reduce the probability of successful attacks~\cite{zhu2016differential}.
Similarly, attribute unlearning methods also adopt the PoT setting~\cite{li2023making}. After model training is completed, attribute unlearning protects private attributes by minimizing the differences between embeddings.
%
%
However, existing methods have three main issues: \textit{Issue 1. Insufficient recommendation preservation.} As shown in Figure~\ref{fig:diff}, due to limited accessibility, PoT approaches cannot directly utilize training data, leading to a noticeable decline in the model's recommendation performance;
\textit{Issue 2. Lack of theoretical guarantees.} Existing methods are based on intuition guidance, lacking a theoretical guarantee of defense; 
and \textit{Issue 3. Instability of adversarial training.} Although the adversarial training methods extend defense from PoT to In-Training (InT) settings, the instability of the optimization process makes it difficult to ensure convergence.

\textbf{Our work.} To address these issues, we propose a privacy--preserving recommender system named \model{} (Recommendation Attribute Inference Defender). 
%
Regarding \textit{Issue 1}, we extend the defense method to the InT setting, thereby fully utilizing the training data to better preserve recommendation performance. In other words, we can fine-tune the model with the original recommendation objective using training data.
%
Regarding \textit{Issue 2}, we define a defensive objective based on the principle that the predictive performance of a classification model (i.e., attacker) on distribution-consistent data is equivalent to random prediction. This means the objective is to minimize the discrepancy among different attribute distributions, consolidating them into a single consistent distribution. To achieve this, we introduce the Wasserstein distance~\cite{givens1984class} as a measure of distribution discrepancy, transforming the AIA defense problem into one where all attribute distributions converge toward a centroid distribution. Our proposed approach provides theoretical guarantees for AIA defense from the perspective of the attacker.
%
%
%
%
Regarding \textit{Issue 3}, we separate the original recommendation objective training with the computation of the centroid distribution (i.e., our proposed additional defensive objective) to ensure more stable optimization. To alleviate potential conflicts between the two objectives, we further impose a recommendation constraint on the centroid distribution, which smooths the optimization trajectory. Additionally, due to the high computational complexity associated with calculating the centroid distribution, we opt to update it at intervals rather than after every training iteration.
%

The main contributions of this paper are:
\begin{itemize}[leftmargin=*] \setlength{\itemsep}{-\itemsep}
    \item We explore defenses against AIA in recommender systems and propose a novel method named \model{}. 
    Our proposed method improves upon traditional attribute unlearning methods and extends them to the InT setting, effectively protecting users' private attributes while preserving recommendation performance.
    
    \item In addition to the original recommendation objective, we define the defense objective against AIA as minimizing the discrepancy in distributions under different attributes. 
    This definition ensures that the training data obtained by the attacker follows a single consistent distribution, thereby rendering the predictive performance of attackers trained on such data equivalent to random prediction.
    
    \item To ensure stability in the training process and improve training efficiency, we separate the recommendation objective training from the computation of constraint centroid distribution, updating the centroid distribution at regular intervals.
    
    \item We conduct extensive experiments and in-depth analyses across mainstream recommendation models on real-world datasets, to evaluate our proposed method w.r.t. privacy protection and recommendation performance. 
    %
\end{itemize}

%% file: section/2-Background.tex
\section{Related Work}

\subsection{Recommender Systems}
\label{sec:related}
Recommender systems analyze user-item interactions and other side information to rank items based on user preferences. 
Among the multitude of recommendation models, CF has been widely applied in practical settings. 
%
The concept of CF is recommending similar items to users with similar behaviors~\cite{koren2009collaborative,shi2014collaborative}.
%
%
%
In CF, let $\mathcal{U}$ and $\mathcal{V}$ represent the sets of users and items, respectively, and $\mathcal{R}$ denotes the user-item interaction behavior matrix. 
The primary objective is to learn the user embedding matrix $P$ and the item embedding matrix $Q$.
In the existing literature~\cite{kluver2018rating}, CF can be categorized into three types:
i) \textbf{Matrix Factorization-based CF}~\cite{mnih2007probabilistic}, which decomposes the user-item interaction matrix $\mathcal{R}$ into the product of two low-dimensional matrices, $P$ and $Q$. 
The product of these two matrices generates a full-ranking matrix to predict the interactions between all users and items;
ii) \textbf{Neural Network-based CF}~\cite{he2017neural}, which encodes each user or item into a latent vector space;
and iii) \textbf{Graph-based CF}~\cite{he2020lightgcn}, which considers the user-item interactions from the perspective of a user-item interaction graph. 
Here, a user's interaction history is akin to the user’s first-order connections, and embeddings of users and items are learned on the graph network.
Regardless of the approach used, the primary objective of CF is to learn embeddings for users and items. 
Thus, our defensive method operates on these embeddings, making it adaptable to most recommendation models.

\subsection{Attribute Inference Attacks}

Recent research demonstrates that users are highly vulnerable to AIA~\cite{otterbacher2010inferring,weinsberg2012blurme,wu2020joint,zheleva2009join,chaabane2012you,gong2016you,gong2018attribute,michalevsky2015powerspy,narain2016inferring,zhang2012cross,fredrikson2014privacy,fredrikson2015model}.
In such attacks, the attacker acquires a dataset related to the target user, referred to as user profiles, to infer the user's private attributes, such as gender, location, political views, or sexual orientation. 
Specifically, the attacker employs a machine learning classifier, inputting the user profiles to predict the user's attribute values. 
This classifier is trained on a shadow dataset containing user profiles alongside corresponding attribute values. 
Such attacks are extensively explored in the following three types of scenarios.

\textbf{Scenario I.} In recommender systems, a user's user profiles might consist of their ratings for certain items. 
Previous study~\cite{weinsberg2012blurme} verifies that attackers can use a logistic regression classifier to predict a user's gender based on their ratings of movies. 
%
%
%
%
%
Subsequent studies~\cite{conti-wb-jayaraman2022attribute,mehnaz2022your} replace the classifier with more effective Multilayer Perceptrons (MLP) to achieve improved results.
\textbf{Scenario II.} In social media contexts~\cite{zheleva2009join,chaabane2012you,gong2016you,gong2018attribute}, attackers can similarly use machine learning classifiers to infer target users' private attributes based on their user profiles on social media, such as pages liked or shared, music playlists, and friends lists. 
The attack pattern in social media closely mirrors that in recommender systems. 
\textbf{Scenario III.} In mobile applications, a previous study~\cite{michalevsky2015powerspy} demonstrates attackers could use classifiers to infer a user's location based on the total power consumption of the user’s smartphone.

The attack patterns across these three scenarios are strikingly similar, with attackers collecting datasets from users who have disclosed their attributes and using this data as a shadow dataset to train classifiers. 
Our defensive methods against AIA in recommender systems described in this paper can be readily extended to other scenarios.

\subsection{AIA Defenses}

To defend against AIA, researchers propose privacy-preserving methods based on various strategies, generally categorized into three types: DP-based methods, attribute unlearning methods, and adversarial training methods. 

\begin{itemize}[leftmargin=*] \setlength{\itemsep}{-\itemsep}
    \item \textbf{DP.} DP aims to add noise to user profiles to prevent an attacker's classifier from accurately inferring users' private attributes~\cite{balu2016differentially,zhu2016differential}. 
    %
    %
    %
    Since additional noise must be introduced into the training data or model parameters, DP methods may reduce the utility of the model~\cite{jia2018attriguard}.
    \item \textbf{Attribute unlearning.} In recommender systems, attribute unlearning methods aim to make the model unlearn certain private attributes of users by training the model under the PoT setting. 
    Specifically, by reducing the distances between sample embeddings during the PoT process, it becomes challenging for attackers to effectively classify the data~\cite{li2023making}. 
    However, since PoT adjustments cannot utilize the original recommendation data, this may significantly reduce the model's recommendation performance.
    \item \textbf{Adversarial training.} 
    %
    %
    %
    %
    Existing research views attackers and recommender systems as adversarial parties, implementing defenses by solving a min-max problem.
    Specifically, Beigi~et~al. introduce a minimax game between Bayesian personalized ranking and worst-case attackers, where user representations can be combined with interaction data to recover sensitive user information~\cite{beigi2020privacy}. 
    Ganhoor~et~al. integrate this approach into a VAE-based recommendation model to remove private information from latent variables~\cite{ganhor2022unlearning}. 
    Although these adversarial training methods are computationally feasible, their training process remains unstable, and they cannot guarantee theoretical privacy protection.
\end{itemize}

%% file: section/3-Preliminary.tex
\section{Preliminary} \label{sec:pre}
To enhance readability, we briefly introduce the notation used in this paper, followed by a succinct overview of the Wasserstein distance and the threat model.

\subsection{Notations} \label{subsec:notation}


\paragraph{Definition of recommendation.} CF involves encoding users and items into embeddings and optimizing these embeddings by minimizing loss functions. 
Let $\mathcal{U} = \{u_1,\ldots,u_{N} \}$ and $\mathcal{V} = \{v_1,\ldots,v_M \}$ respectively represent sets of users and items, with $N$ and $M$ denoting the numbers of users and items, respectively. 
The user-item interaction matrix $\mathcal{R} \in \mathbb{R}^{N\times M}$
denotes the interaction records between each user and item. 
In this paper, we adopt the implicit feedback setting where the element of interaction matrix $r_{u,v}=1$ if there is an interaction between user $u$ and item $v$ else $r_{u,v}=0$.
Generally, many existing CF approaches utilize encoder networks $f(\text{·})$ to generate embeddings of users and items $f(u),f(v) \in \mathbb{R}^d$.

\paragraph{Definition of defense.} Let $\mathcal{D} = \{ (u_n,y_n,a_n) \}_{n=1}^{N}$ as a set of $N$ independent and identically distributed (i.i.d.) samples drawn from an unknown probability distribution over $\mathcal{U} \times \mathcal{Y} \times \mathcal{A}$, where $u_n$ denotes the $n$-th user, with $y_n = f(u_n)$ denotes user embeddings, and $a_n$ represents private attributes. 
Each sample point $(u_n,y_n,a_n)$ corresponds to a user profile record. 
Given the number attribute classes $K$, we denote $\mathcal{D}^i = \{ (u_n,y_n,a_n) \in \mathcal{D}, \text{ } s.t. \text{ } a_n = a^i, i \in [1,2,\ldots,K] \}$ as the set of users have the attribute $a^i$, where $a^i$ represents the $i$-th category of attribute $a$.
The number of users in this set is denoted as $|\mathcal{D}^i| = N^i$, and $y_n \in \mathcal{D}^i$ is also denoted as $y_n^i$.
Note that this paper focuses on protecting a single privacy attribute, with one defensive objective corresponding to each attribute. 
The term ``attribute distribution'' refers to the embedding distribution of users with the same class label in the target attribute. 
The target attribute is the one to unlearn, and attribute labels denote specific categories (e.g., ``male'' or ``female'').
Depending on the context, we use either ``class distribution'' or ``attribute distribution'', but they refer to the same concept in this paper. 
Additionally, while users may have multiple privacy attributes that require protection, our proposed framework can be readily adapted to protect multiple attributes simultaneously by incorporating multiple defensive objectives.
Let $U$, $Y$, and $A$ be random variables corresponding to $x$, $u$, and $a$ respectively, each having a probability density function $p(\text{·})$. 
Thus, define $Y^i$ as the random variable relative to set $\mathcal{D}^i$, we have the probability density function $p(Y^i) = p(Y \mid A=a^i)$. 

\subsection{Wasserstein Distance}

In this paper, we employ Wasserstein distance~\cite{givens1984class} as the distance measurement between distributions in subsequent derivations. To enhance the readability, we briefly introduce its definition in this section.

Given two probability measures $\alpha$ and $\beta$ on spaces $\mathcal{X}$ and $\mathcal{Y}$ respectively, each endowed with densities $p_{X}$ and $p_{Y}$ (i.e., $d \alpha(x) = p_{X}(x) dx$ and $d \beta(y) = p_{Y}(y) dy$), 
consider a transportation map $T: \mathcal{X} \to \mathcal{Y}$ that for any measurable subset $\mathcal{B} \in \mathcal{Y}$ satisfies $\int_{\mathcal{B}} p_{Y} (y)dy = \int_{T^{-1}(\mathcal{B})} p_{X}(x) dx$ (mass conservation constraint).
This means that the mass of the measurable subset $\mathcal{B}$ w.r.t. the density $p_{Y}$ is equal to the mass of the set $T^{-1}(\mathcal{B})$ w.r.t. the density $p_{X}$. 
Let $\mathcal{T}$ be the set of all transportation maps from $\mathcal{X}$ to $\mathcal{Y}$, and let $c:\mathcal{X} \times \mathcal{Y} \to [0,+\infty]$ be a cost function where $c(x,T(x))$  represents the cost of transporting $x$ to $T(x)$. 
In the original formulation~\cite{monge1781memoire}, the optimal transportation map $T$ is the map that minimizes the total transportation cost:
\begin{equation}
    \min_{T} \left\{\int_{\mathcal{X}} c(x, T(x)) \mathrm{d} \alpha(x): T_{\sharp} \alpha=\beta \right\},
    \label{eq:monge}
\end{equation}
%
%
where $T_{\sharp}$ represents the push-forward operator that satisfies the mass conservation constraint when mapping the probability measure $\alpha$ onto $\beta$.

Given the non-convex nature of the feasible set in the aforementioned problem, Kantorovich~\cite{kantorovich1942transfer} reformulates the optimal transport problem by finding the optimal joint distribution in the set of joint distributions on $\mathcal{X} \times \mathcal{Y}$, denoted as 
\begin{equation}
    \min_{T} \left\{\int_{\mathcal{X} \times \mathcal{Y}} c(x, y) \mathrm{d} T(x, y) : P_{\mathcal{X}\sharp} T = \alpha,  P_{\mathcal{Y}\sharp} T = \beta \right\},
    \label{eq:kanto}
\end{equation}
where $P_{\mathcal{X}\sharp}$ and $P_{\mathcal{Y}\sharp}$ are the push-forward operators of the projection $P_{\mathcal{X}} (x,y)=x$ and $P_{\mathcal{Y}} (x,y)=y$, which map the joint distribution to the respective marginal distributions. 
Additionally, $T$ has a density $p_{X \times Y}$ such that $d T(x,y) = p_{X \times Y} dxdy$. 
This reformulation allows the problem to be addressed within a convex framework, facilitating the application of linear programming techniques to find solutions that minimize the transportation cost across the distributions.

Assuming that $\mathcal{X} = \mathcal{Y}$ and for some $p \geq 1$ such that $c(x, y)=d(x, y)^{p}$, where $d(x, y)$ is the distance on space $\mathcal{X}$, the $p$-Wasserstein distance is defined as
\begin{equation}
    \mathcal{W}_{p} (\alpha, \beta) = \min_{T} \left( \int_{\mathcal{X} \times \mathcal{Y}} c(x, y) \mathrm{d} T(x, y) \right)^{\frac{1}{p}},
    \label{eq:wp}
\end{equation}
where $T$ must satisfy the same constraints in Eq. (\ref{eq:kanto}).
$\mathcal{W}_{p}$ is symmetric and positive, $\mathcal{W}_{p} (\alpha, \beta) = 0$ if and only if $\alpha = \beta$, and it satisfies the triangle inequality
\begin{equation}
    \forall \alpha, \beta, \gamma \in \mathcal{S},  \mathcal{W}_{p} (\alpha, \beta) \leq \mathcal{W}_{p} (\alpha, \gamma) + \mathcal{W}_{p} (\gamma, \beta),
    \label{eq:wp-inequality}
\end{equation}
where $\mathcal{S}$ is the space containing $\alpha$, $\beta$, and $\gamma$.

\subsection{Threat Model} \label{subsec:threat}


\paragraph{Adversary’s goal}

The goal of the attacker is to infer users' sensitive attributes from available information. 
%
%
%
In this paper, we focus on discrete attributes with two or more labels. 
Assuming the information possessed by the attacker can be represented as 
$I=\{(y_n,a_n)\}_{n=1}^{\hat{N}}$, where $\hat{N} < N$ denotes the number of users with available attributes leaked to the attacker. 
The attacker aims to train a classifier $\vartheta(\text{·}): \mathbb{R}^{d} \to \{1,2,\ldots,K \}$.

\paragraph{Adversary’s knowledge} \label{subsec:know}

In recommender systems, the knowledge relied upon by AIA can generally be categorized into two types: 
\textit{i) Model parameters.} Existing research~\cite{ganhor2022unlearning,li2023making} indicates that basic machine learning models, due to their representative learning capabilities, can successfully infer user attributes from user embeddings learned through CF models. 
%
%
%
\textit{ii) Recommendation results.} Furthermore, personalized recommendation results can also be exploited by attackers, as they significantly reflect user preferences and attribute information~\cite{weinsberg2012blurme}. 
%
%
Following prior studies~\cite{ganhor2022unlearning,li2023making}, we assume a gray-box scenario where attackers access partial model parameters, i.e., user embeddings.
The adversary is able to access an IID shadow dataset of available information above.


%% file: section/4-Method.tex
\section{Method}

\begin{figure*}[t]
\centering
\includegraphics[width=0.95\linewidth]{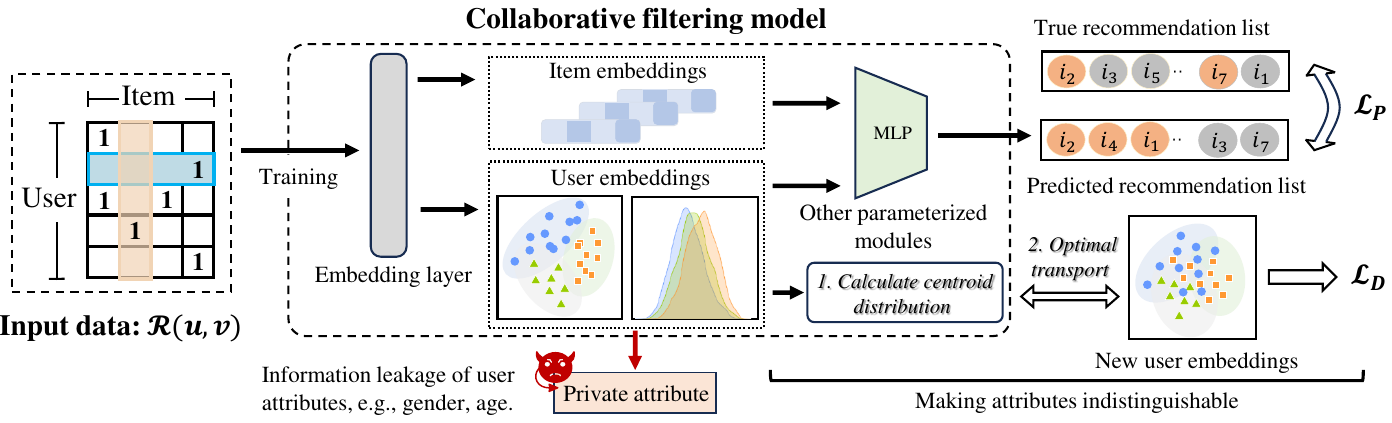} 
\caption{
%
%
%
\model{}, established under the InT setting, can directly use the original training data to preserve recommendation performance. 
Additionally, \model{} defines a defensive objective to ensure that the class distribution of protected attributes is independent of class labels, which is equivalent to making all class distributions indistinguishable. 
Based on this, \model{} first calculates a centroid distribution that satisfies the constraints on recommendation performance and then uses optimal transport to align all class distributions with it, rendering the class distributions indistinguishable.}
\label{fig:pipeline}
\end{figure*}

In this section, we introduce our proposed privacy-preserving recommender system, \model{}. 
%
%
An overview of \model{} is shown in Figure~\ref{fig:pipeline}.
We first introduce the definitions of both the recommendation and defensive objectives, then present the training algorithm that integrates these two objectives.

\subsection{In-training for Recommendation}

\paragraph{Recommendation objective} 
The recommendation model acquires embeddings for users and items, 
and performs recommendation tasks by utilizing the similarity (e.g., vector dot product) between user and item embeddings. 
As described in Section~\ref{subsec:notation}, most CF models learn embeddings for users and items through an encoder network $f(\text{·})$.
For instance, neural network-based models typically employ neural network layers as encoders, while graph-based models incorporate neighborhood information into the encoder.

To preserve recommendation performance, \model{} is constructed under InT setting, allowing the use of model's training data to enhance recommendation performance. 
%
%
Specifically, following prior work~\cite{he2017neural}, we adopt the widely-used Cross-Entropy (CE) loss as the recommendation objective:
\begin{equation}
    \mathcal{L}_{P} = \frac{\sum_{(u,v) \in \mathcal{R} \cup \mathcal{R}^{-}} \log \left( \left( s_{u,v} \right)^{r_{u,v}} \left(1 - s_{u,v} \right)^{(1-r_{u,v})} \right)} {|\mathcal{R} \cup \mathcal{R}^{-}|} ,
    \label{eq:ce}
\end{equation}
%
%
where $\mathcal{R}^{-} = \{ (u,v^{-})\}$ 
represents the set of uninteracted negative samples, and $s_{u,v}$ denotes the predicted rating of interaction $(u,v)$.
%
%
For all recommendation models described in this paper, we employ the CE loss during training.

\subsection{Defending Against AIA}

\paragraph{Defensive objective} Since the attacker is trained on the user embedding by the recommendation model, defenders (i.e., from the perspective of recommender systems) can intervene in the attacker's training dataset (i.e., user embeddings) to degrade the effectiveness of the classifier model trained by the attacker. 
Given that the attacker is typically unknown, the defense strategy proposed in this paper is based on the attacker's training dataset.
Specifically, the defender's objective can be expressed as 
\begin{equation}
    \mathcal{P}(Y \mid \boldsymbol{A}=\boldsymbol{a}^i)=\mathcal{P}(Y \mid \boldsymbol{A}=\boldsymbol{a}^j), \quad \forall i \neq j, \boldsymbol{a}^i, \boldsymbol{a}^j \in \mathcal{A}.
    \label{eq:defend_obj_origin}
\end{equation}
This indicates that the user embeddings obtained under different classes by the attacker have the same distribution, implying a minimal correlation between the training data and private user attributes.
Consequently, a classifier trained on such data will perform no better than random prediction.

Let $\mathcal{P}(Y^i)=\mathcal{P}(Y \mid A=a^i)$. 
In order to achieve the target expressed in Eq. (\ref{eq:defend_obj_origin}), we minimize the following loss function:
\begin{equation}
    \mathcal{L}_{D}(\theta) = \min_{\theta} \sum_{i=1}^{K} \sum_{\substack{j=1,j \neq i}}^{K} Dis(\mathcal{P}(Y^i) , \mathcal{P}(Y^j)),
    \label{eq:defend_dis}
\end{equation}
where $\mathcal{P}(Y^i)$ denotes the output distribution for the $i$-th category and $Dis(\text{·} , \text{·})$ represents the distance between two distributions.
In this paper, we employ the squared 2-Wasserstein distance ($\mathcal{W}_2$ distance) to measure the distance between distributions. 
The motivation for using the $\mathcal{W}_2$ distance lies in its well-defined and computable nature, even between distributions with non-overlapping supports.
Consequently, Eq. (\ref{eq:defend_dis}) can be reformulated as follows: 
\begin{equation}
    \mathcal{L}_{D}(\theta) = \min_{\theta} \sum_{i=1}^{K} \sum_{\substack{j=1,j \neq i}}^{K} \mathcal{W}_2^2(\mathcal{P}(Y^i), \mathcal{P}(Y^j)).
    \label{eq:eq:defend_dis_w}
\end{equation}
Given that the Wasserstein distance satisfies the triangle inequality~(\ref{eq:wp-inequality}), it yields an upper bound for Eq. (\ref{eq:eq:defend_dis_w}):
%
\begin{align}
    \mathcal{L}_{D}(\theta)
    & \leq \min_{\theta} \sum_{i=1}^{K} \sum_{\substack{j=1,j \neq i}}^{K}\left[\mathcal{W}_2(\mathcal{P}(Y^i), \mathcal{P}^*) + \mathcal{W}_2(\mathcal{P}(Y^j), \mathcal{P}^*))\right]^2 \notag \\
    & \leq \min_{\theta} \sum_{i=1}^{K} \sum_{\substack{j=1,j \neq i}}^{K}2\left[\mathcal{W}_2^2(\mathcal{P}(Y^i), \mathcal{P}^*) + \mathcal{W}_2^2(\mathcal{P}(Y^j), \mathcal{P}^*))\right] \notag \\
    & = \min_{\theta} 4(K-1) \sum_{i=1}^{K} \mathcal{W}_2^2(\mathcal{P}(Y^i), \mathcal{P}^*),
    \label{eq:defend_upper_bound}
\end{align}
\normalsize
where $\mathcal{P}^*$ is an unknown distribution located in the same space as $\mathcal{P}(Y^i)$.
Upon obtaining $\mathcal{P}^*$, Eq. (\ref{eq:defend_upper_bound}) can be directly optimized using gradient descent, a process equivalent to making all output distributions $\{ \mathcal{P}(Y^i) \}$ converge towards $\mathcal{P}^*$.
Therefore, based on this upper bound, we can minimize Eq. (\ref{eq:eq:defend_dis_w})  by minimizing Eq. (\ref{eq:defend_upper_bound}).

\begin{figure}[t]
\centering
\includegraphics[width=1\columnwidth]{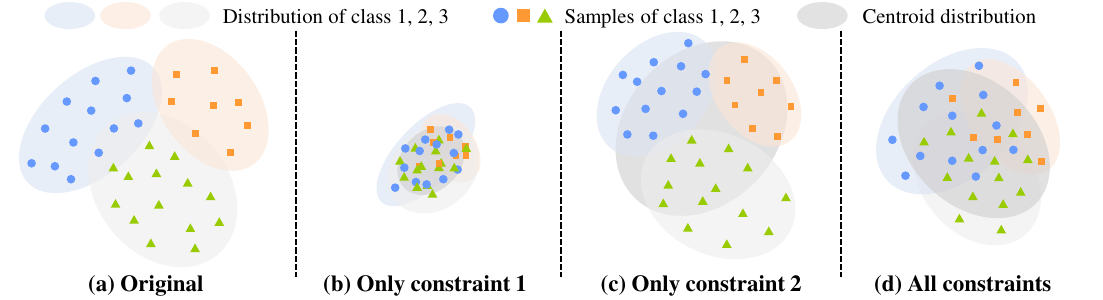} 
\caption{The function of each constraint. (a) In the initial stage, distinct decision boundaries between class distributions provide potential attacking foundations. (b) When Constraint 1 is applied, distributions are correctly centered but too compact, leading to significant overlap among user embeddings and reducing their ability to distinguish information beyond protected attributes. (c) When Constraint 2 is applied, the centroids become overly relaxed and unstable, causing only slight movements of each class distribution. There remains a clear decision boundary between the class distributions. 
}
\label{fig:subject}
\end{figure}

\paragraph{Restrictions on Distribution $\mathcal{P}^*$} Although aligning each class distribution closely with distribution $\mathcal{P}^*$ can achieve the goal of defense, not every distribution $\mathcal{P}^*$ is suitable for preserving recommendation performance. 
Thus, we set the following two constraints on distribution $\mathcal{P}^*$: 
\begin{itemize}[leftmargin=*] \setlength{\itemsep}{-\itemsep}
    \item \textit{Constraint 1.} According to~\cite{han2024intra}, to minimize the impact on recommendation performance, class distributions with a larger number of samples should be adjusted less, i.e., be closer to $\mathcal{P}^*$, while those with fewer samples should be adjusted more, i.e., be further from $\mathcal{P}^*$.
    \item \textit{Constraint 2.} Distribution $\mathcal{P}^*$ should not be overly compact. If $\mathcal{P}^*$ is too compact, user embeddings may collapse to this overly compact distribution, not only losing the information of the protected attributes but also potentially losing other important information. 
    %
\end{itemize}
To more clearly convey our ideas, we illustrate the role of each constraint in Figure~\ref{fig:subject}.
To provide a formalized representation, we first define the differential entropy:
\begin{equation}
    H(\mathcal{P}) := -\int_X \log \left( \frac{d\mathcal{P}}{dx} \right) d\mathcal{P}(x),
    \label{eq:entropy}
\end{equation}
where a smaller $H(\mathcal{P})$ implies that the distribution $\mathcal{P}$ is more compact.
Subsequently, these two constraints can be expressed as
\begin{align}
    \mathcal{P}^* &= \underbrace{\argmin_{\mathcal{P}} \sum_{i=1}^{K} \lambda_{i} \mathcal{W}_2^2(\mathcal{P}(Y^i), \mathcal{P})}_{Constraint\enspace 1} \underbrace{- \tau H(\mathcal{P})}_{Constraint\enspace 2}, 
    \label{eq:wb}
\end{align}
where $\lambda_i$ represents the weight of class distribution $\mathcal{P}(Y^i)$, equivalent to the size-normalized summation~\cite{agueh2011barycenters}, 
and $\tau \geq 0$ is a penalty coefficient used to control the preference for the entropy penalty term $-H(\mathcal{P})$.
$\mathcal{P}^*$ is called penalized 2-Wasserstein barycenter.

\paragraph{Empirical computation} In practice, since constructing the centroid from the overall distribution $\{ \mathcal{P}(Y^i) \}$ is infeasible, we empirically compute the centroid distribution $\mathcal{P}^*$ using the empirical distribution $\hat{\mathcal{P}}(Y^i)$ obtained from $\mathcal{D}^i$. 
Following~\cite{klenke2013probability,jiang2020wasserstein}, if the samples in $\mathcal{D}$ are i.i.d., as $|\mathcal{D}| \rightarrow \infty$, with $\mathcal{W}_{2}^{2} \left(\mathcal{P}(Y^i), \mathcal{P}^* \right)<\infty$ for all $i \in [1,2,\ldots,K]$, the empirical barycenter distribution with penalty term satisfies $\lim \sum_{i=1}^K \mathcal{W}_{2}^{2} (\hat{\mathcal{P}}(Y^i), \hat{\mathcal{P}}^* ) \rightarrow \sum_{i=1}^K \mathcal{W}_{2}^{2} \left(\mathcal{P}(Y^i), \mathcal{P}^* \right)$ almost surely. 
%
%
Subsequently, all computations related to the distribution can be executed empirically.
%
%

\subsection{Putting Together}

This section puts together our proposed objectives and formally introduces our \model{} framework along with the corresponding computational implementation.

\paragraph{Overview of \model{}} We define the total loss function over dataset $\mathcal{D} = \{ (u_n,y_n,a_n) \}_{n=1}^{\hat{N}}$ as 
\begin{equation}
    \mathcal{L}_{ALL}(\theta) = \mathcal{L}_{P}({\theta}) + \eta \sum_{i=1}^{K} \mathcal{W}_{2}^{2} \left( \hat{\mathcal{P}}(Y^i), \hat{\mathcal{P}}^* \right),
    \label{eq:empirical}
\end{equation}
where $\eta$ is an adjustable coefficient and the last term is the loss of defensive objective $\mathcal{L}_{D}(\theta)$.
These two terms represent the losses of recommendation objective and defensive objective, respectively. 
However, in the initial stage of training, the user embeddings do not adequately reflect their category characteristics, meaning that the centroid distribution computed at this time lacks significance, thus rendering the latter term of the equation less effective. 
Therefore, \model{} adopts a two-phase training strategy: 
\begin{itemize}[leftmargin=*] \setlength{\itemsep}{-\itemsep}
    \item \textbf{Phase I.} In the initial stage, we only optimize the first term of Eq. (\ref{eq:empirical}) to ensure recommendation performance. 
    \item \textbf{Phase II.} Once the model achieves preliminary fitting, we optimize the entire equation.
    Given that the latter term cannot be directly calculated, it must be approached in two steps: 
    \textit{Step 1.} Calculating the empirical centroid distribution $\mathcal{P}^*$; 
    \textit{Step 2.} Treating it as a constant independent of model parameters, and optimizing the entire Eq. (\ref{eq:empirical}) using gradient descent.
\end{itemize}

\paragraph{Computational implementation} Since the loss term in Phase I equals the original recommendation model training, we focus on addressing the computational issues in Phase II, which are divided into two steps.


For \textit{step 1}, we utilize following Lemma~\ref{lem:com_wb} to compute the empirical barycenter $\hat{\mathcal{P}}^*$ on $\{ \bar{y}_q \}_{q=1}^{\bar{N}}$, where $\{ \bar{y}_q \}_{q=1}^{\bar{N}}$ represents the atoms (support set) of the centroid distribution $\hat{\mathcal{P}}^*$, where $\delta_y$ is the Dirac distribution at $y$.
The empirical wasserstein barycenter can be estimated by $\hat{\mathcal{P}}^* =  \sum_{n=1}^{\hat{N}} \alpha_n \delta_{\bar{y}_n}, \ \sum_{n=1}^{\hat{N}} \alpha_n =1$.

%
%
\begin{lemma}
    Given sample-based empirical distributions $\hat{\mathcal{P}}(Y^i) = \frac{1}{N_i}\sum_{n=1}^{N_i} \delta_{y_n^i}, i\in [K]$, the dual problem of penalized Wasserstein barycenter can be formulated as
    %
    \begin{align}
    \label{dual-problem}
    \max_{g_1,\cdots,g_K}\sum_{i=1}^K \lambda_i \sum_{n=1}^{N_i} g_i^{c}(y_n^i) - \tau \log  \sum_{n=1}^{\hat{N}} \exp \left( -\frac{1}{\tau} \sum_{i=1}^{K} \lambda_i g_i(\bar{y}_n) \right),
    \end{align}
    \normalsize
    where $g(\text{·})$ denotes the dual problem and $g^{c}_i(y_n^i)$ is defined by $$\min_{p=1,\cdots,\hat{N}} \{ \| y_n^i - \bar{y}_p \|^2 - g_i(\bar{y}_p)\}.$$ Moreover, Eq.~(\ref{dual-problem}) is concave and L-Lipschitz continuous with respect to $g_i$'s. If ${(g_i)}^n_{i=1}$ solves the dual problem, then $(g_i^c, g_i)_{i=1}^{K}$ can be utilized to recover the empirical wasserstrin barycenter $\hat{\mathcal{P}}^*$ and the optimal couplings $\{T_i^*\}_{i=1}^K$.
    \label{lem:com_wb}
\end{lemma}

\begin{proof}
    The proof is given in Appendix~\ref{app:proofs}.
\end{proof}

For \textit{step 2}, to compute the gradient of Eq. (\ref{eq:empirical}), it is necessary to solve for the optimal coupling $T_i^*$ between each class distribution $\hat{\mathcal{P}}(Y^i)$ and the centroid distribution $\hat{\mathcal{P}}^*$.
Since we have computed $\mathcal{P}^*$ and $\{ \bar{y}_n \}_{n=1}^{\bar{N}}$. 
The calculation of the optimal coupling family $\{ T_{i}^{*} \}$ can be solved by iterative KL-projection method~\cite{benamou2015iterative}.
Based on the optimal coupling $\{T_i^*\}$, the gradient of Eq. (\ref{eq:empirical}) w.r.t. $\theta$ can be further derived by the following Lemma: 
\begin{lemma}
    Assuming datasets $\{y^{i}_{p}\}_{p=1}^{N^{i}}$ and $\{\bar{y}_q \}_{q=1}^{\bar{N}}$ correspond to empirical distributions $\hat{\mathcal{P}}(Y^i)$ and $\hat{\mathcal{P}}^*$ respectively, both distributions $\hat{\mathcal{P}}(Y^i)$ and $\hat{\mathcal{P}}^*$ exist within the same space $\mathcal{S}=\mathbb{R}^d$. 
    If $c(y^i_p,\bar{y}_q)=\| y^i_p - \bar{y}_q\|^2$, we can derive the gradient of $\mathcal{L}_{D}(\theta)$ satisfies:
    %
    \begin{equation}
        \nabla_{\theta} \mathcal{L}_{D}(\theta) = 2\sum_{i=1}^K \sum_{p=1}^{N_i}{\left[ \partial y_p^i(\theta) \right]}^{\top} \left( \frac{y_p^i}{N^i} - \sum_{q=1}^{\bar{N}} T_{i}^{*}(p,q) \bar{y}_q \right)_{p=1}^{N^i},
    \notag
    \end{equation}
    \normalsize
    where $\partial y_p^i(\theta) \in \mathbb{R}^{dim(\theta) \times (N^i d)}$ is the Jacobian of the map $y_p^i(\theta)$ and $T_{i}^{*}(p,q)=T_{i}^{*}(y_p^i,\bar{y}_q)$ represents the transfer amount from $y_p^i$ to $\bar{y}_q$ in the optimal coupling matrix $T_{i}^{*}$.
    \label{lem:com_g}
\end{lemma}

\begin{proof}
    The proof is given in Appendix~\ref{app:proofs}.
\end{proof}

Since the gradient of loss $L_{P}(\theta)$ can be directly computed, based on Lemma~\ref{lem:com_g}, the gradient for Eq.~(\ref{eq:empirical}) can be determined with $\nabla_{\theta} \mathcal{L}_{ALL}(\theta) = \nabla_{\theta} \mathcal{L}_{P}(\theta) + \eta \nabla_{\theta} \mathcal{L}_{D}(\theta)$. Subsequent updates of parameters adhere to $\theta_{t+1} = \theta_{t} - \mu \nabla_{\theta} \mathcal{L}_{ALL}(\theta)$, where $\mu$ represents the step size. 
According to previous studies~\cite{cuturi2013sinkhorn,deshpande2018generative,jiang2020wasserstein}, under the condition of optimal coupling $T_{i}^{*}$ being sparse, the computation of $\nabla_{\theta} \mathcal{L}_{D}(\theta)$ can be completed in linear time.
However, it is important to note that calculating the centroid distribution $\hat{\mathcal{P}}^*$ entails a significant computational cost. 
Therefore, we do not compute it at every step; instead, we opt to perform this calculation every $\xi$ step, where $\xi$ is dynamically adjusted based on the progress of training. For a detailed description of the algorithmic process, please refer to Algorithm~\ref{alg:\model{}-framework}.

\begin{algorithm}[t]
\caption{\model{} Framework}
\label{alg:\model{}-framework}
\begin{algorithmic}[1] 
\State \textbf{Inputs:} Dataset $\mathcal{D} = \{ (u_n,y_n,a_n) \}_{n=1}^{N}$, adjustable coefficients $\eta$, update step size $\mu$, optimization rounds $E^1$ of Phase I, optimization rounds $E^2$ of Phase II, frequency of barycenter distribution computation $\xi$.
\State \textbf{Initialize:} Model parameters $\theta_0$.
\For{$i = 1$ \textbf{to} $E^1$} \Comment{Phase I}
    \State Compute gradients $\nabla \mathcal{L}_{P}(\theta)$.
    \State Update parameters $\theta_{i} = \theta_{i-1} - \mu \nabla \mathcal{L}_{P}(\theta)$.
\EndFor
\For{$i = E^1+1$ \textbf{to} $(E^1+E^2)$} \Comment{Phase II}
    \If{$\left((i-E^1)\mod \xi \right) == 0$} 
        \State Compute the empirical barycenter distribution $\hat{\mathcal{P}}^*$
        \Statex \hspace{\algorithmicindent}\hspace{\algorithmicindent} with penalty term and support set $\{ \bar{y}_n \}_{n=1}^{\hat{N}}$ as
        \Statex \hspace{\algorithmicindent}\hspace{\algorithmicindent} defined in Lemma~\ref{lem:com_wb}.
    \EndIf
    \State Compute gradients $\nabla \mathcal{L}_{D}(\theta)$ as defined in Lemma~\ref{lem:com_g}.
    \State Update parameters $\theta_{i} = \theta_{i-1} - \mu \nabla \mathcal{L}_{ALL}(\theta)$.
\EndFor
\State \Return $\theta_{E^1+E^2}$.
\end{algorithmic}
\end{algorithm}

%% file: section/5-Evaluation.tex
\section{Evaluation}
In this section, we evaluate \model{} across four real-world datasets, focusing on performance of defense, recommendation, and efficiency. Additionally, we conduct an ablation study to comprehensively analyze the role of the proposed loss components.

\subsection{Experimental Settings}

\begin{table}[t]
\centering
\caption{Summary of datasets.}
\label{tab:dataset}
\resizebox{\linewidth}{!}{
\begin{tabular}{lcccccc}
\toprule
Dataset & Attribute & Category \# & User \#  & Item \#   & Rating \# & Sparsity \\
\midrule
\multirow{2}{*}{ML-1M}   
& Gender & 2 & \multirow{2}{*}{6,040} & \multirow{2}{*}{3,416} & \multirow{2}{*}{999,611} & \multirow{2}{*}{95.155\%} \\
& Age & 3 & & & &\\
\midrule
\multirow{2}{*}{LFM-2B}  
& Gender & 2 & \multirow{2}{*}{19,972} & \multirow{2}{*}{99,639}   & \multirow{2}{*}{2,829,503}  & \multirow{2}{*}{99.858\%} \\
& Age & 3 & & & & \\
\midrule
\multirow{2}{*}{KuaiSAR} 
& Feat1 & 7 & \multirow{2}{*}{21,852} & \multirow{2}{*}{140,367}   & \multirow{2}{*}{2,166,893}  & \multirow{2}{*}{99.929\%} \\
& Feat2 & 2 & & & \\
\midrule
\multirow{2}{*}{Last.fm-360K} 
& Gender & 2 & \multirow{2}{*}{130,046} & \multirow{2}{*}{11,835} & \multirow{2}{*}{6,450,050} & \multirow{2}{*}{99.581\%} \\
& Age & 3 & & & & \\
\bottomrule
\end{tabular}
}
\end{table}

\paragraph{Datasets} The experiments are conducted on four real-world public datasets, each comprising user-item interaction data and user attribute data. 
The user attribute data includes sensitive information such as gender, age, and location. 
Detailed information about the datasets is as follows:
\begin{itemize}[leftmargin=*] \setlength{\itemsep}{-\itemsep}
    \item 
    \textbf{MovieLens 1M (ML-1M)}: A version of MovieLens dataset that has 1 million ratings.
    \item 
    \textbf{LFM-2B}\footnote{http://www.cp.jku.at/datasets/LFM-2b}: This dataset collected more than 2 billion listening events, which are used for music retrieval and recommendation tasks~\cite{melchiorre2021investigating}. 
    %
    Here, we use a subset of the whole dataset which includes more than 3 million ratings.
    \item
    \textbf{KuaiSAR}\footnote{https://kuaisar.github.io/}: KuaiSAR is a unified search and recommendation dataset containing user behavior logs collected from the short-video mobile app, Kuaishou\footnote{https://www.kuaishou.com/}. Here we use a tiny version of KuaiSAR, i.e., KuaiSAR-small. 
    \item 
    \textbf{Last.fm-360K}\footnote{http://ocelma.net/MusicRecommendationDataset/lastfm-360K.html}: This dataset contains approximately 17 million listening events from nearly 360,000 users, and is commonly used for music retrieval and recommendation tasks~\cite{Celma:Springer2010}.
\end{itemize}
%
%
The characteristics are summarized in Table~\ref{tab:dataset}.
%


\paragraph{Recommendation models} We validate the effectiveness of our method on three mainstream recommendation models. 
These models represent three common structural types mentioned in Section~\ref{sec:related}, including Matrix Factorization-based CF, Neural Network-based CF, and Graph-based CF.
\begin{itemize}[leftmargin=*] \setlength{\itemsep}{-\itemsep}
    \item 
    \textbf{DMF}: Deep Matrix Factorization~\cite{xue2017deep} is one of the representative models based on matrix factorization.
    \item 
    \textbf{NCF}: Neural Collaborative Filtering~\cite{he2017neural} is a classical CF model based on neural networks.
    \item 
    \textbf{LightGCN}: Light Graph Convolution Network~\cite{he2020lightgcn} is the State-Of-The-Art (SOTA) CF model that improves recommendation performance by simplifying graph convolution network. 
\end{itemize}

\paragraph{Evaluation Metrics} We evaluate the performance of defense and recommendation using the following metrics.

\begin{itemize}[leftmargin=*] \setlength{\itemsep}{-\itemsep}
\item \textbf{Defense.} 
Aligning consistently with~\cite{li2023making}, attackers can train a robust adversarial classifier with machine learning models. 
To quantify the effectiveness of the defense, we utilize two commonly used classification metrics: the micro-averaged F1 score (F1) and Balanced Accuracy (BAcc) to assess the performance of AIA~\cite{grandini2020metrics,ganhor2022unlearning}. 
Lower values of F1 and BAcc indicate a more effective defense. 
According to existing literature~\cite {beigi2020privacy,zhang2021graph}, we train the attacker using 80\% of the users, with the remaining 20\% used for testing. 
The outcomes of the AIA are obtained through five-fold cross-validation, averaged over five runs. 
To ensure a fair comparison, we optimized the hyperparameters until the loss function converged to obtain the optimal defensive effect.

\item \textbf{Recommendation.} To assess the recommendation performance, we employ the leave-one-out testing~\cite{he2017neural}. 
We utilize Hit Rate at rank $K$ (HR@$K$) and Normalized Discounted Cumulative Gain at rank $K$ (NDCG@$K$) as recommendation metrics. 
HR@$K$ measures whether the test item appears in the top-$K$ list, while NDCG@$K$ is a position-aware ranking metric that assigns higher scores to hits at higher positions~\cite{he2015trirank,xue2017deep}. 
In this paper, we calculate HR@$K$ and NDCG@$K$ using the entire set of negative items rather than a sampled subset, due to observed instability and inconsistency issues with sampled metrics compared to their precise versions~\cite{krichene2020sampled}. 
Note that we compare the recommendation performance of various methods when achieving optimal defense effects separately.
\end{itemize}

\paragraph{Compared baselines} Although the setting of PoT differs from that of InT, comparing our proposed InT methods with PoT approaches contributes to a comprehensive understanding of the problem. 
Therefore, we compare our proposed methods with three representative baselines:
\begin{itemize}[leftmargin=*] \setlength{\itemsep}{-\itemsep}
    \item \textbf{Original}: This is the original model without defense.
    \item \textbf{DP} (DP-PoT): This method protects user attributes by introducing noise perturbations to the original training data~\cite{zhu2016differential}. Under the gray-box setting in this paper, DP is implemented by adding noise to embedding layer parameters.
    \item \textbf{AU} (AU-PoT): This represents the current SOTA PoT method~\cite{li2023making}. It optimizes a two-component loss. 
    \item \textbf{Adv} (Adv-InT): This method uses adversarial training to achieve attribute defense for the MultVAE~\cite{ganhor2022unlearning}. We also apply the idea of adversarial training to our tested recommendation models, i.e., DMF, NCF, and LightGCN.
\end{itemize}

\paragraph{Hardware Information} All models and algorithms are implemented with Python 3.8 and PyTorch 1.9. We run all experiments on an Ubuntu 20.04 System server with 256GB RAM and NVIDIA GeForce RTX 4090 GPU.

We run all models 10 times and report the average results.
We provide details of dataset pre-processing and parameter settings in Appendix~\ref{sec:exp_details}.

\subsection{Results} \label{sucsec:results}

\paragraph{Defense performance} Protecting the target attributes is our primary focus. 
The performance of the defense is assessed by the performance of the attacker, i.e., an MLP. 
%
%
From the results in Table~\ref{tab:defense}, we make the following observations. 
Firstly, the attacker achieves an average F1 score of 0.5322 and BAcc of 0.5320 on the original user embeddings, indicating that user attribute information embedded in user embeddings can be inferred by the attacker. 
Secondly, all methods can protect the attribute information contained in user embeddings to varying degrees. 
The methods of DP-PoT, AU-PoT, Adv-InT, and RAID reduce the F1 score by an average of 16.68\%, 25.30\%, 15.92\%, and 41.35\%, respectively. 
At the same time, AU-PoT, Adv-InT, and RAID can reduce the BAcc by an average of 21.57\%, 12.96\%, and 28.85\%, respectively. 
In contrast, DP-PoT only reduces BAcc by 19.15\%. 
For binary attributes such as gender and multi-class attributes such as age, the BAcc of attackers using the RAID method post-defense is equivalent to that of a random attacker. 
This indicates that our proposed RAID method can effectively protect private attributes in recommendation models, whether they are binary or multi-class.

%

\begin{table}[t]
\centering
\caption{Performance of attribute inference attack. The top results are highlighted in \textbf{bold}.}
\label{tab:defense}
\resizebox{\linewidth}{!}{
\begin{tabular}{lcccccccc}
\toprule
\multirow{2}{*}{Dataset} & \multirow{2}{*}{Attribute} & \multirow{2}{*}{Method} & \multicolumn{2}{c}{DMF} & \multicolumn{2}{c}{NCF} & \multicolumn{2}{c}{LightGCN} \\ 
\cmidrule{4-5} \cmidrule{6-7} \cmidrule{8-9}
& & & F1 & BAcc & F1 & BAcc & F1 & BAcc \\
\midrule


\multirow{10}{*}{ML-1M} & \multirow{5}{*}{Gender} 
& Original & 0.7501 & 0.7487 & 0.7545 & 0.7485 & 0.6987 & 0.7023 \\
&  & DP-PoT & 0.4425 & \textbf{0.4425} & 0.7318 & 0.7341 & 0.4290 & \textbf{0.4290} \\
&  & AU-PoT & 0.4868 & 0.4910 & 0.2889 & 0.5000 & 0.4047 & 0.5016 \\
&  & Adv-InT & 0.6180 & 0.6106 & 03042 & 0.5182 & 0.5626 & 0.5980 \\
&  & RAID & \textbf{0.2666} & 0.5000 & \textbf{0.2665} & \textbf{0.4999} & \textbf{0.2667} & 0.5000 \\ \cmidrule{2-9}
& \multirow{5}{*}{Age} & Original & 0.6221 & 0.6222 & 0.6241 & 0.6241 & 0.5664 & 0.5664 \\
&  & DP-PoT & 0.4254 & 0.4255 & 0.6109 & 0.6109 & 0.4155 & 0.4155 \\
&  & AU-PoT & 0.3336 & \textbf{0.3333} & \textbf{0.3151} & 0.4998 & 0.3333 & \textbf{0.3333} \\
&  & Adv-InT & 0.5994 & 0.5995 & 0.4524 & 0.4524 & 0.3923 & 0.3923 \\
&  & RAID & \textbf{0.3334} & \textbf{0.3333} & 0.3832 & \textbf{0.3834} & \textbf{0.3331} & \textbf{0.3333} \\
\midrule

\multirow{10}{*}{LFM-2B} & \multirow{5}{*}{Gender} & Original & 0.6717 & 0.6802 & 0.6808 & 0.6779 & 0.6219 & 0.6162 \\
&  & DP-PoT & 0.5552 & 0.5559 & 0.6727 & 0.6691 & 0.3336 & \textbf{0.3336} \\
&  & AU-PoT & 0.5109 & 0.5088 & 0.4526 & \textbf{0.4469} & 0.4816 & 0.4701 \\
&  & Adv-InT & 0.6336 & 0.6259 & 0.6129 & 0.6163 & 0.5704 & 0.5733 \\
&  & RAID & \textbf{0.3999} & \textbf{0.5000} & \textbf{0.2667} & 0.5000 & \textbf{0.3110} & 0.5000 \\ \cmidrule{2-9}
& \multirow{5}{*}{Age} & Original & 0.3376 & 0.3378 & 0.3334 & 0.3341 & 0.3362 & 0.3362 \\
&  & DP-PoT & \textbf{0.3333} & 0.3338 & 0.3328 & 0.3324 & 0.3333 & 0.3332 \\
&  & AU-PoT & \textbf{0.3333} & \textbf{0.3333} & \textbf{0.3160} & \textbf{0.3159} & 0.3325 & 0.3323 \\
&  & Adv-InT & 0.3338 & \textbf{0.3333} & 0.3333 & 0.3333 & 0.3206 & 0.3207 \\
&  & RAID & \textbf{0.3333} & \textbf{0.3333} & 0.3317 & 0.3333 & \textbf{0.3185} & \textbf{0.3193} \\
\midrule

\multirow{10}{*}{KuaiSAR} & \multirow{5}{*}{Feat1} & Original & 0.2530 & 0.2373 & 0.1522 & 0.1500 & 0.2160 & 0.2132 \\
&  & DP-PoT & 0.1603 & 0.1569 & 0.1549 & 0.1590 & 0.1042 & 0.1250 \\
&  & AU-PoT & 0.1275 & 0.1375 & 0.1203 & 0.1250 & 0.1249 & 0.1222 \\
&  & Adv-InT & 0.1182 & 0.1417 & 0.10665 & 0.1278 & 0.1531 & 0.1556 \\
&  & RAID & \textbf{0.1042} & \textbf{0.1250} & \textbf{0.0949} & \textbf{0.1125} & \textbf{0.0972} & \textbf{0.1167} \\ \cmidrule{2-9}
& \multirow{5}{*}{Feat2} & Original & 0.5685 & 0.5685 & 0.4242 & 0.4333 & 0.5564 & 0.5833 \\
&  & DP-PoT & 0.4394 & 0.4444 & 0.5503 & 0.5777 & \textbf{0.2782} & \textbf{0.3333} \\
&  & AU-PoT & 0.5490 & 0.5544 & \textbf{0.1309} & \textbf{0.1311} & 0.3558 & 0.3630 \\
&  & Adv-InT & 0.4006 & 0.4037 & 0.2848 & 0.3407 & \textbf{0.2782} & \textbf{0.3333} \\
&  & RAID & \textbf{0.2782} & \textbf{0.3333} & 0.2781 & 0.3333 & \textbf{0.2782} & \textbf{0.3333} \\
\midrule

\multirow{10}[4]{*}{Last.fm-360K} & \multirow{5}[2]{*}{Gender} & Original & 0.7017  & 0.7119  & 0.7073  & 0.7159  & 0.7300  & 0.7310  \\
&  & DP-PoT & 0.6721  & 0.6825  & 0.6731  & 0.6782  & 0.6672  & 0.6691  \\
&  & AU-PoT & 0.7015  & 0.7055  & 0.6983  & 0.7005  & 0.7228  & 0.7246  \\
&  & Adv-InT & 0.4922  & 0.5052  & 0.9821  & 0.9822  & 0.7818  & 0.7787  \\
&  & RAID  & \textbf{0.4857 } & \textbf{0.4862 } & \textbf{0.3111 } & \textbf{0.5000 } & \textbf{0.6242 } & \textbf{0.6438 } \\ \cmidrule{2-9}          
& \multirow{5}[2]{*}{Age} & Original & 0.5658  & 0.5658  & 0.5373  & 0.5373  & 0.5789  & 0.5789  \\
&  & DP-PoT & 0.4621  & 0.4657  & 0.5061  & 0.5061  & 0.5146  & 0.5146  \\
&  & AU-PoT & 0.5394  & 0.5394  & 0.4855  & 0.4855  & 0.5580  & 0.5580  \\
&  & Adv-InT & \textbf{0.4077 } & \textbf{0.4077 } & 0.6315  & 0.6315  & 0.5514  & 0.5514  \\
&  & RAID  & 0.4356  & 0.4356  & \textbf{0.3333 } & \textbf{0.3333 } & \textbf{0.4865 } & \textbf{0.4762 } \\
\bottomrule
\end{tabular}
}
\end{table}

\begin{table*}[t]
\centering
\caption{Results of recommendation performance with truncating the ranking list for these metrics at 5 and 10. The top results are highlighted in \textbf{bold}.}
\label{tab:rec_performance_5_10}
\resizebox{\linewidth}{!}{
\begin{tabular}{lcccccccccccccc}
\toprule
\multirow{2}{*}{Dataset} & \multirow{2}{*}{Attribute} & \multirow{2}{*}{Method} & \multicolumn{4}{c}{DMF} & \multicolumn{4}{c}{NCF} & \multicolumn{4}{c}{LightGCN} \\ 
\cmidrule(lr){4-7} \cmidrule(lr){8-11} \cmidrule(lr){12-15}
& & & HR@5 & NDCG@5 & HR@10 & NDCG@10 & HR@5 & NDCG@5 & HR@10 & NDCG@10 & HR@5 & NDCG@5 & HR@10 & NDCG@10 \\
\midrule


\multirow{10}{*}{ML-1M} & \multirow{5}{*}{Gender} 
& Original & 0.0630 & 0.0390 & 0.1040 & 0.0520 & 0.0660 & 0.0410 & 0.1060 & 0.0540 & 0.0600 & 0.0370 & 0.1000 & 0.0500 \\
&  & DP-PoT & 0.0290 & 0.0180 & 0.0490 & 0.0240 & 0.0090 & 0.0050 & 0.0150 & 0.0070 & 0.0260 & 0.0170 & 0.0420 & 0.0220 \\
&  & AU-PoT & \textbf{0.0600} & \textbf{0.0380} & \textbf{0.1000} & \textbf{0.0510} & 0.0480 & 0.0320 & 0.0800 & 0.0420 & 0.0500 & 0.0320 & 0.0800 & 0.0410 \\
&  & Adv-InT & 0.0560 & 0.0350 & 0.0970 & 0.0490 & 0.0530 & 0.0320 & 0.0980 & 0.0470 & 0.0510 & 0.0320 & 0.0830 & 0.0430 \\
&  & RAID & 0.0540 & 0.0340 & 0.0890 & 0.0460 & \textbf{0.0530} & \textbf{0.0330} & \textbf{0.0980} & \textbf{0.0470} & \textbf{0.0550} & \textbf{0.0340} & \textbf{0.0850} & \textbf{0.0440} \\ \cmidrule{2-15}
& \multirow{5}{*}{Age} 
& Original & 0.0630 & 0.0390 & 0.1040 & 0.0520 & 0.0660 & 0.0410 & 0.1060 & 0.0540 & 0.0600 & 0.0370 & 0.1000 & 0.0500 \\
&  & DP-PoT & 0.0220 & 0.0130 & 0.0440 & 0.0200 & 0.0090 & 0.0050 & 0.0150 & 0.0070 & 0.0250 & 0.0160 & 0.0430 & 0.0210 \\
&  & AU-PoT & 0.0500 & 0.0330 & 0.0850 & 0.0430 & 0.0570 & 0.0370 & 0.0950 & 0.0490 & 0.0510 & 0.0320 & 0.0790 & 0.0410 \\
&  & Adv-InT & 0.0510 & 0.0330 & 0.0880 & 0.0450 & 0.0550 & 0.0340 & 0.0990 & 0.0480 & 0.0450 & 0.0290 & 0.0740 & 0.0380 \\
&  & RAID & \textbf{0.0540} & \textbf{0.0340} & \textbf{0.0880} & \textbf{0.0450} & \textbf{0.0590} & \textbf{0.0380} & \textbf{0.1010} & \textbf{0.0520} & \textbf{0.0550} & \textbf{0.0340} & \textbf{0.0890} & \textbf{0.0450} \\
\midrule

\multirow{10}{*}{LFM-2B} & \multirow{5}{*}{Gender} 
& Original & 0.0120 & 0.0070 & 0.0200 & 0.0100 & 0.0170 & 0.0100 & 0.0280 & 0.0140 & 0.0220 & 0.0140 & 0.0340 & 0.0180 \\
&  & DP-PoT & 0.0030 & 0.0010 & 0.0050 & 0.0020 & 0.0010 & 0.0010 & 0.0020 & 0.0010 & 0.0040 & 0.0020 & 0.0060 & 0.0030 \\
&  & AU-PoT & 0.0050 & 0.0030 & 0.0080 & 0.0040 & 0.0110 & 0.0070 & 0.0200 & 0.0100 & \textbf{0.0210} & \textbf{0.0130} & \textbf{0.0320} & \textbf{0.0170} \\
&  & Adv-InT & 0.0080 & 0.0050 & 0.0160 & 0.0070 & 0.0120 & 0.0070 & 0.0210 & 0.0100 & 0.0190 & 0.0110 & 0.0300 & 0.0150 \\
&  & RAID & \textbf{0.0090} & \textbf{0.0050} & \textbf{0.0180} & \textbf{0.0080} & \textbf{0.0140} & \textbf{0.0090} & \textbf{0.0240} & \textbf{0.0120} & 0.0190 & 0.0120 & 0.0310 & 0.0160 \\ \cmidrule{2-15}
& \multirow{5}{*}{Age} 
& Original & 0.0120 & 0.0070 & 0.0200 & 0.0100 & 0.0170 & 0.0100 & 0.0280 & 0.0140 & 0.0220 & 0.0140 & 0.0340 & 0.0180 \\
&  & DP-PoT & 0.0020 & 0.0010 & 0.0050 & 0.0020 & 0.0010 & 0.0010 & 0.0020 & 0.0010 & 0.0030 & 0.0020 & 0.0050 & 0.0030 \\
&  & AU-PoT & 0.0100 & 0.0060 & 0.0150 & 0.0080 & 0.0110 & 0.0070 & 0.0200 & 0.0100 & 0.0200 & \textbf{0.0130} & \textbf{0.0320} & \textbf{0.0170} \\
&  & Adv-InT & 0.0080 & 0.0050 & 0.0160 & 0.0080 & 0.0130 & 0.0080 & 0.0220 & 0.0110 & 0.0160 & 0.0100 & 0.0280 & 0.0140 \\
&  & RAID & \textbf{0.0110} & \textbf{0.0070} & \textbf{0.0170} & \textbf{0.0090} & \textbf{0.0170} & \textbf{0.0110} & \textbf{0.0260} & \textbf{0.0140} & \textbf{0.0200} & 0.0120 & 0.0310 & 0.0150 \\
\midrule

\multirow{10}{*}{KuaiSAR} & \multirow{5}{*}{Feat1} 
& Original & 0.0190 & 0.0120 & 0.0310 & 0.0150 & 0.0180 & 0.0110 & 0.0320 & 0.0150 & 0.0200 & 0.0130 & 0.0360 & 0.0180 \\
&  & DP-PoT & 0.0040 & 0.0020 & 0.0080 & 0.0040 & 0.0020 & 0.0010 & 0.0050 & 0.0020 & 0.0060 & 0.0040 & 0.0110 & 0.0050 \\
&  & AU-PoT & 0.0150 & 0.0100 & 0.0270 & 0.0130 & 0.0180 & 0.0110 & 0.0300 & 0.0150 & 0.0190 & 0.0120 & 0.0320 & 0.0160 \\
&  & Adv-InT & 0.0190 & 0.0120 & 0.0310 & 0.0150 & 0.0150 & 0.0100 & 0.0260 & 0.0140 & 0.0120 & 0.0080 & 0.0210 & 0.0100 \\
&  & RAID & \textbf{0.0190} & \textbf{0.0120} & \textbf{0.0320} & \textbf{0.0160} & \textbf{0.0160} & \textbf{0.0110} & \textbf{0.0300} & \textbf{0.0150} & \textbf{0.0190} & \textbf{0.0120} & \textbf{0.0330} & \textbf{0.0160} \\ \cmidrule{2-15}
& \multirow{5}{*}{Feat2} 
& Original & 0.0170 & 0.0100 & 0.0300 & 0.0150 & 0.0180 & 0.0110 & 0.0320 & 0.0150 & 0.0200 & 0.0130 & 0.0360 & 0.0180 \\
&  & DP-PoT & 0.0030 & 0.0020 & 0.0050 & 0.0020 & 0.0020 & 0.0010 & 0.0050 & 0.0020 & 0.0050 & 0.0030 & 0.0100 & 0.0050 \\
&  & AU-PoT & 0.0140 & 0.0090 & 0.0270 & 0.0130 & 0.0160 & \textbf{0.0110} & \textbf{0.0300} & \textbf{0.0150} & 0.0160 & 0.0100 & 0.0290 & 0.0140 \\
&  & Adv-InT & 0.0090 & 0.0050 & 0.0190 & 0.0080 & 0.0080 & 0.0050 & 0.0160 & 0.0080 & 0.0110 & 0.0070 & 0.0180 & 0.0090 \\
&  & RAID & \textbf{0.0160} & \textbf{0.0090} & \textbf{0.0290} & \textbf{0.0130} & \textbf{0.0170} & 0.0100 & 0.0290 & 0.0140 & \textbf{0.0190} & \textbf{0.0120} & \textbf{0.0320} & \textbf{0.0160} \\
\midrule

\multirow{10}[4]{*}{last.FM 360K} & \multirow{5}[2]{*}{Gender} & Original & 0.0630  & 0.0400  & 0.1050  & 0.0530  & 0.0630  & 0.0400  & 0.1070  & 0.0540  & 0.0670  & 0.0420  & 0.1120  & 0.0570  \\
&  & DP-PoT & 0.0520  & 0.0320  & 0.0880  & 0.0440  & 0.0500  & 0.0310  & 0.0870  & 0.0430  & 0.0180  & 0.0110  & 0.0300  & 0.0150  \\
&  & AU-PoT & \textbf{0.0630 } & \textbf{0.0400 } & \textbf{0.1050 } & \textbf{0.0530 } & \textbf{0.0630 } & \textbf{0.0400 } & \textbf{0.1070 } & \textbf{0.0540 } & \textbf{0.0670 } & \textbf{0.0420 } & \textbf{0.1120 } & \textbf{0.0570 } \\
&  & Adv-InT & 0.0610  & 0.0380  & 0.1020  & 0.0510  & 0.0620  & 0.0390  & 0.1040  & 0.0520  & 0.0550  & 0.0360  & 0.0910  & 0.0470  \\
&  & RAID  & \textbf{0.0630 } & \textbf{0.0400 } & \textbf{0.1050 } & \textbf{0.0530 } & 0.0620  & \textbf{0.0400 } & 0.1040  & 0.0530  & \textbf{0.0670 } & \textbf{0.0420 } & \textbf{0.1120 } & \textbf{0.0570 } \\
\cmidrule{2-15}          & \multirow{5}[2]{*}{Age} & Original & 0.0630  & 0.0400  & 0.1050  & 0.0530  & 0.0630  & 0.0400  & 0.1070  & 0.0540  & 0.0670  & 0.0420  & 0.1120  & 0.0570  \\
&  & DP-PoT & 0.0520  & 0.0320  & 0.0880  & 0.0440  & 0.0500  & 0.0310  & 0.0870  & 0.0430  & 0.0180  & 0.0110  & 0.0300  & 0.0150  \\
&  & AU-PoT & \textbf{0.0630 } & \textbf{0.0400 } & \textbf{0.1050 } & \textbf{0.0530 } & \textbf{0.0640 } & \textbf{0.0400 } & \textbf{0.1070 } & \textbf{0.0540 } & \textbf{0.0670 } & \textbf{0.0420 } & \textbf{0.1120 } & \textbf{0.0570 } \\
&  & Adv-InT & 0.0600  & 0.0370  & 0.1010  & 0.0500  & 0.0570  & 0.0350  & 0.0980  & 0.0490  & 0.0620  & 0.0400  & 0.1030  & 0.0530  \\
&  & RAID  & \textbf{0.0630 } & \textbf{0.0400 } & \textbf{0.1050 } & \textbf{0.0530 } & 0.0620  & 0.0390  & 0.1060  & 0.0530  & \textbf{0.0670 } & \textbf{0.0420 } & \textbf{0.1120 } & \textbf{0.0570 } \\
\bottomrule
\end{tabular}
}
\end{table*}

\paragraph{Recommodation performance} Recommendation performance is another crucial objective we must consider, as the process of protecting user privacy attributes typically comes at the cost of sacrificing model accuracy. 
To verify the effectiveness of the RAID framework in maintaining recommendation performance, we employ NDCG and HR for evaluation, 
truncating the ranking lists at 5, 10, 15, and 20. 
As shown in Tables~\ref{tab:rec_performance_5_10} and \ref{tab:rec_performance_15_20} (in Appendix B), protecting user attributes against does impact recommendation performance. 
It is also noticeable that methods based on the PoT setting exhibit a significant decline in recommendation performance compared to those based on the InT approach. 
Specifically, compared to the original performance, PoT methods such as DP-PoT and AU-PoT show an average reduction in NDCG of 59.44\% and 8.53\% respectively, and a decrease in HR of 58.46\% and 9.86\% respectively. 
In contrast, the InT method, i.e., Adv-InT, show an average decrease of only 14.76\% in NDCG and 14.73\% in HR. 
RAID, which is also an InT method, records an average decline of 6.01\% in NDCG and 6.44\% in HR, achieving optimal performance across each dataset and model.
Our proposed \model{} surpasses Adv-InT in maintaining recommendation performance because our defense approach, which aligns distributions, does not interfere with other relevant information within user embeddings. 
This preservation of essential embedding features ensures that the recommendation capability remains robust despite the implementation of privacy-preserving measures.





\paragraph{Ablation study} Table~\ref{tab:ablation} presents the results of an ablation study conducted using NCF on the ML-1M dataset. 
We sequentially remove defense loss $\mathcal{L}_{D}$ and recommendation loss $\mathcal{L}_{P}$ to assess their impact on the defense (F1 and BAcc) and recommendation (HR and NDCG) performance. 
Initially, when we remove defense loss $\mathcal{L}_{D}$, we observe a significant increase in the attacker's performance. 
This indicates that without active defense measures, attackers can easily extract private attribute information from user embeddings. 
Subsequently, when we remove recommendation loss $\mathcal{L}_{P}$, although the model performed well in defending against attackers, there is a noticeable decline in recommendation performance. 
This suggests that removing this component of the loss, while allowing the model to remain secure, significantly impairs its recommendation capabilities, decreasing its utility. 
These results from the ablation study clearly demonstrate the critical roles of defense loss $\mathcal{L}_{D}$ and recommendation loss $\mathcal{L}_{P}$ in our proposed RAID. 
The former primarily affects the effectiveness of defense, while the latter ensures the model's utility.

\begin{table}[t]
\centering
\caption{Results of ablation studies.}
\label{tab:ablation}
    \begin{tabular}{cccccc}
    \toprule
$\mathcal{L}_{D}$ & $\mathcal{L}_{P}$ & F1 & BAcc & HR@10 & NDCG@10 \\ \midrule
\ding{51} & \ding{51} & 0.2665 & 0.4999 & 0.0980 & 0.0470 \\
\ding{55} & \ding{51} & 0.7545 & 0.7485 & 0.1060 & 0.0540 \\
\ding{51} & \ding{55} & 0.3973 & 0.5002 & 0.0600 & 0.0290 \\ \bottomrule
    \end{tabular}
\end{table}


%% file: section/6-Discussion.tex
\section{Discussions}

In this section, we discuss the limitations present in existing methods and further validate the effectiveness of our proposed RAID.
Furthermore, we examine the effectiveness of \model{} in defending against attacks targeting multiple attributes and provide a detailed discussion of \model{}'s limitations in Appendix~\ref{sec:dis_app}.

\subsection{Limitation of Existing Methods} \label{subsec:limit}

In response to the limitations of existing methods outlined in Section~\ref{sec:intro}, we conduct individual verifications. 
These evaluations analyze the differences between PoT and InT in their ability to preserve recommendation performance and the effectiveness of defense. 

\begin{table}
\caption{Impacts of PoT and InT settings.}
\label{tab:pot-int}
\centering
\resizebox{\linewidth}{!}{
\begin{tabular}{ccccccc}
\toprule
\multirow{2}{*}{Settings} & \multicolumn{2}{c}{AU-PoT} & \multicolumn{2}{c}{Adv-InT} & \multicolumn{2}{c}{RAID} \\ \cmidrule{2-7}
& HR@10 & NDCG@10 & HR@10 & NDCG@10 & HR@10 & NDCG@10 \\ \midrule
PoT & 0.1530 & 0.0830 & 0.1491 & 0.0720 & 0.1500 & 0.0810 \\
InT & 0.1540 & 0.0835 & 0.1510 & 0.0740 & 0.1550 & 0.0850 \\ \bottomrule
\end{tabular}
}
\end{table}

\paragraph{PoT vs InT} 

To verify the differing impacts of PoT and InT settings on preserving recommendation performance, we attempt to extend them to another setting. 
We control the methods under these two settings to achieve similar defensive performance and compare their effects on preserving recommendation performance. 
Note that DP can only be operated under the PoT setting. 
We validate the performance of model NCF on dataset ML-100K (Gender), and the results are presented in Table~\ref{tab:pot-int}. 
We have the following two observations: 
i) Except for DP, the recommendation performance under the InT setting significantly surpasses that under the PoT setting, confirming our hypothesis;
and ii) Our proposed RAID shows superior recommendation performance in the InT setting compared to the existing methods, indicating that RAID is more compatible with InT.

\paragraph{Effectiveness of the Defense} 

\begin{figure*}[t]
\centering
\subfigure[\textbf{ML-1M (Original)}]{\includegraphics[width=4.2cm]{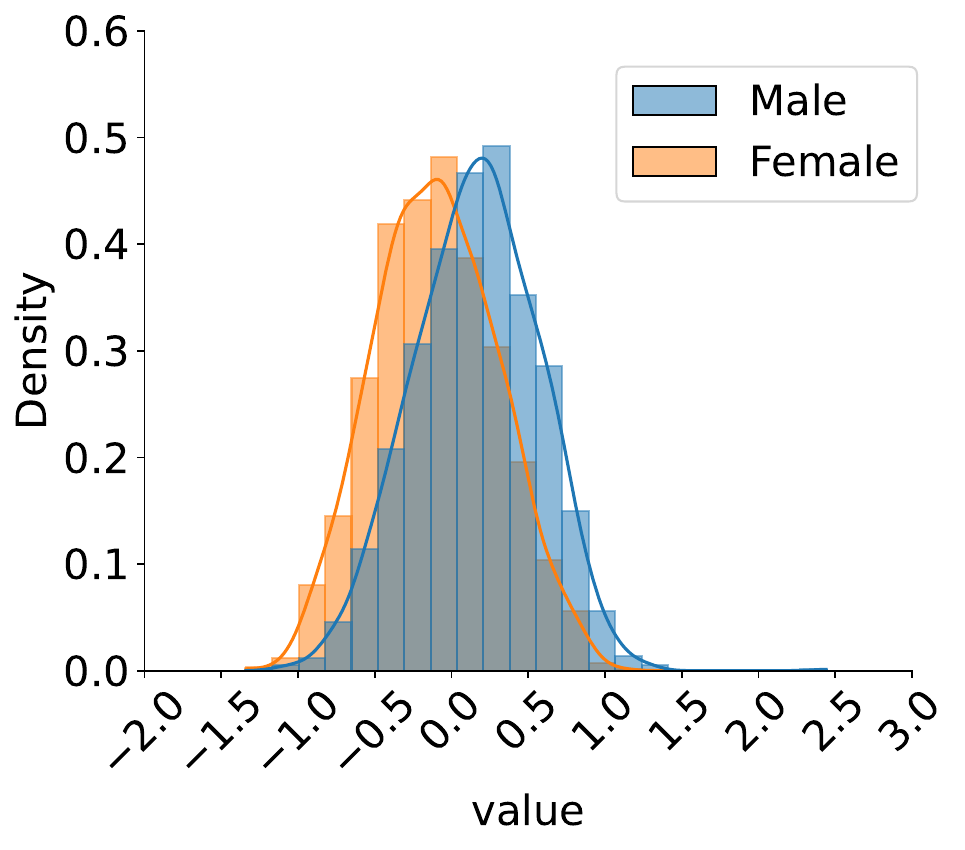}}
\subfigure[\textbf{LFM-2B (Original)}]{\includegraphics[width=4.2cm]{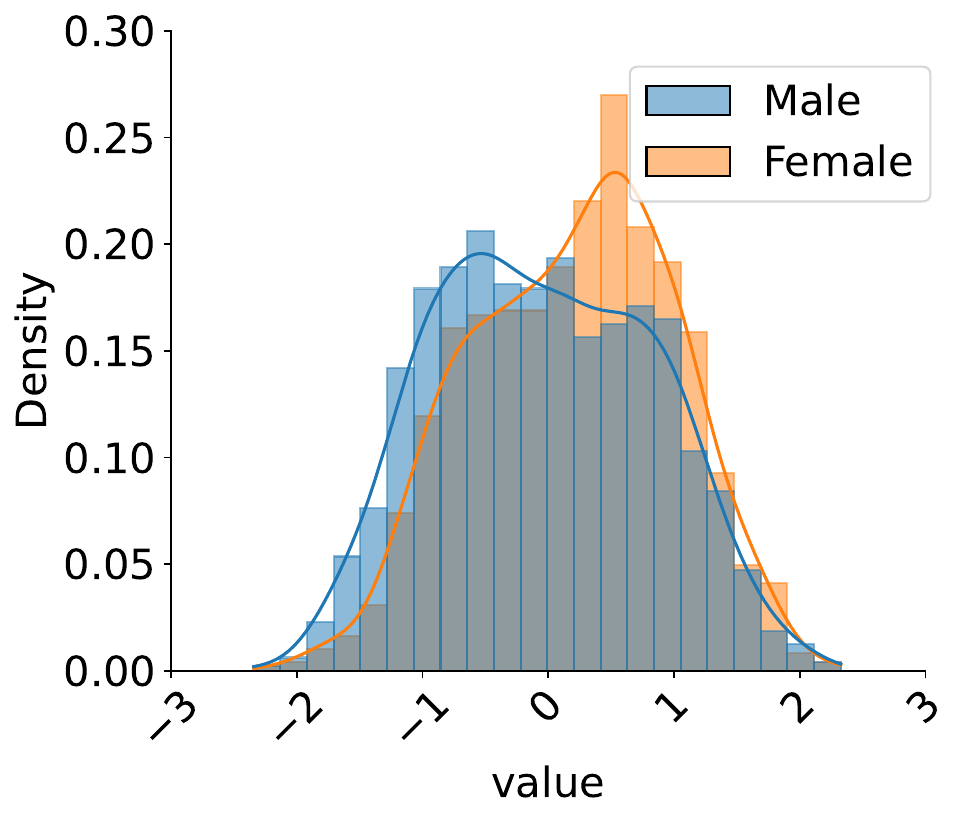}}
\subfigure[\textbf{KuaiSAR (Original)}]{\includegraphics[width=4.2cm]{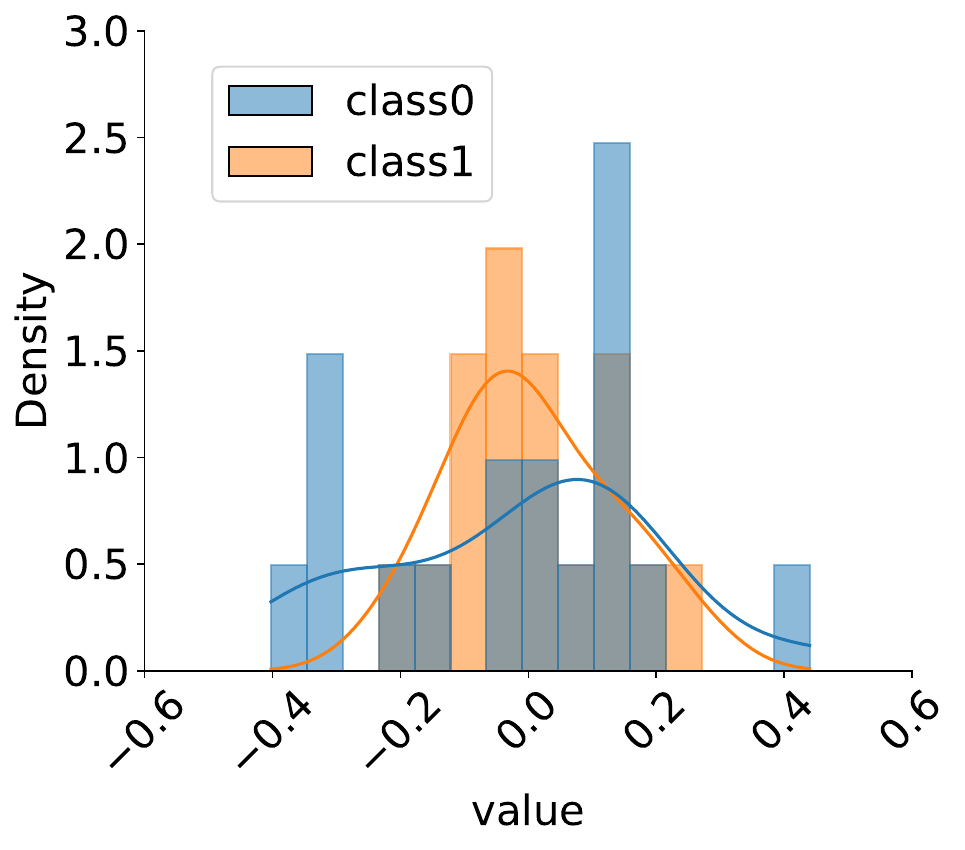}}
\subfigure[\textbf{Last.fm-360K (Original)}]{\includegraphics[width=4.2cm]{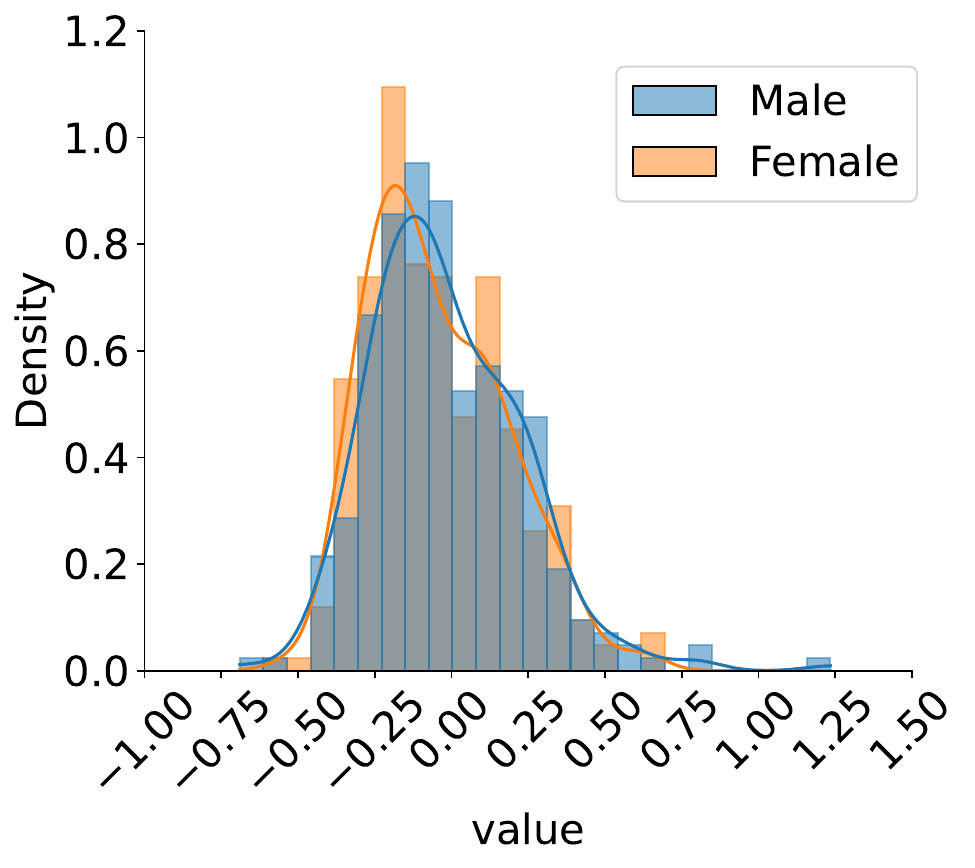}}
\subfigure[\textbf{ML-1M (RAID)}]{\includegraphics[width=4.2cm]{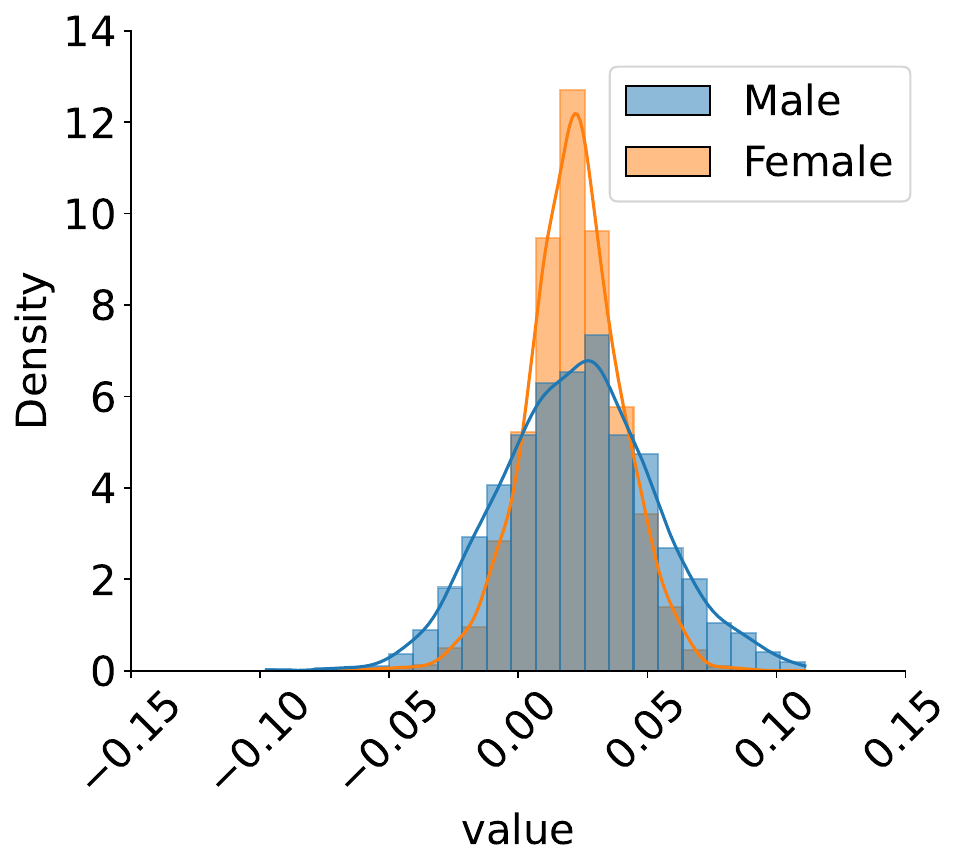}}
\subfigure[\textbf{LFM-2B (RAID)}]{\includegraphics[width=4.2cm]{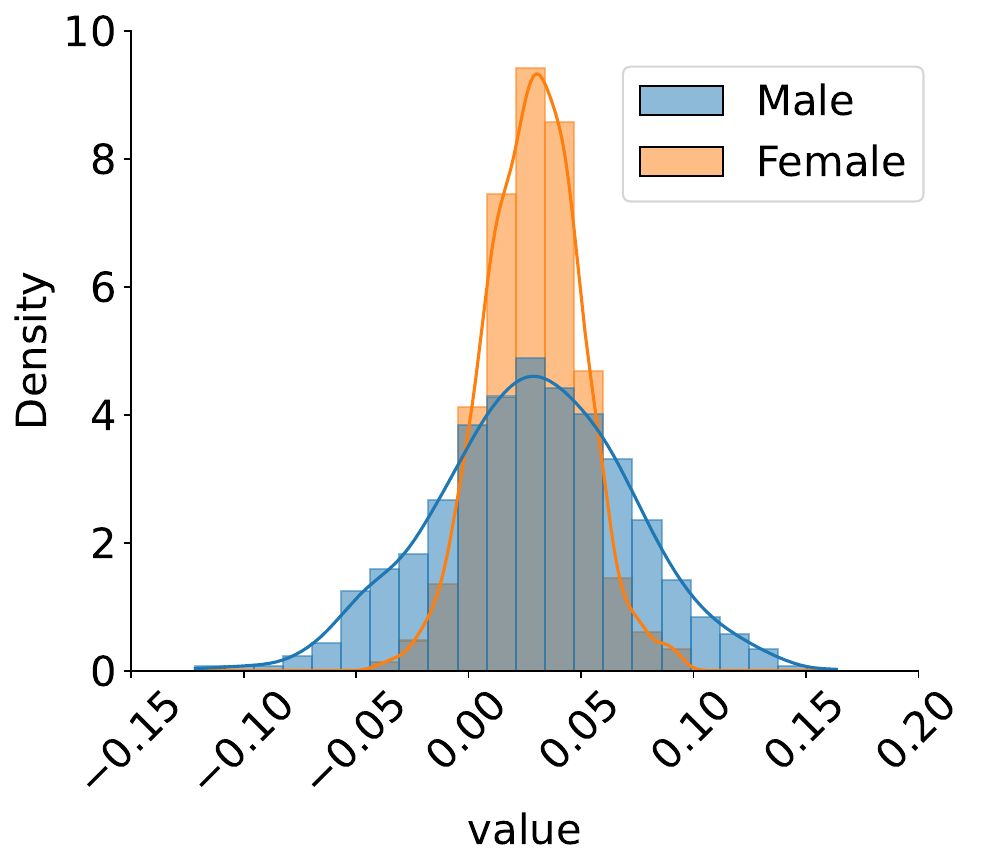}}
\subfigure[\textbf{KuaiSAR (RAID)}]{\includegraphics[width=4.2cm]{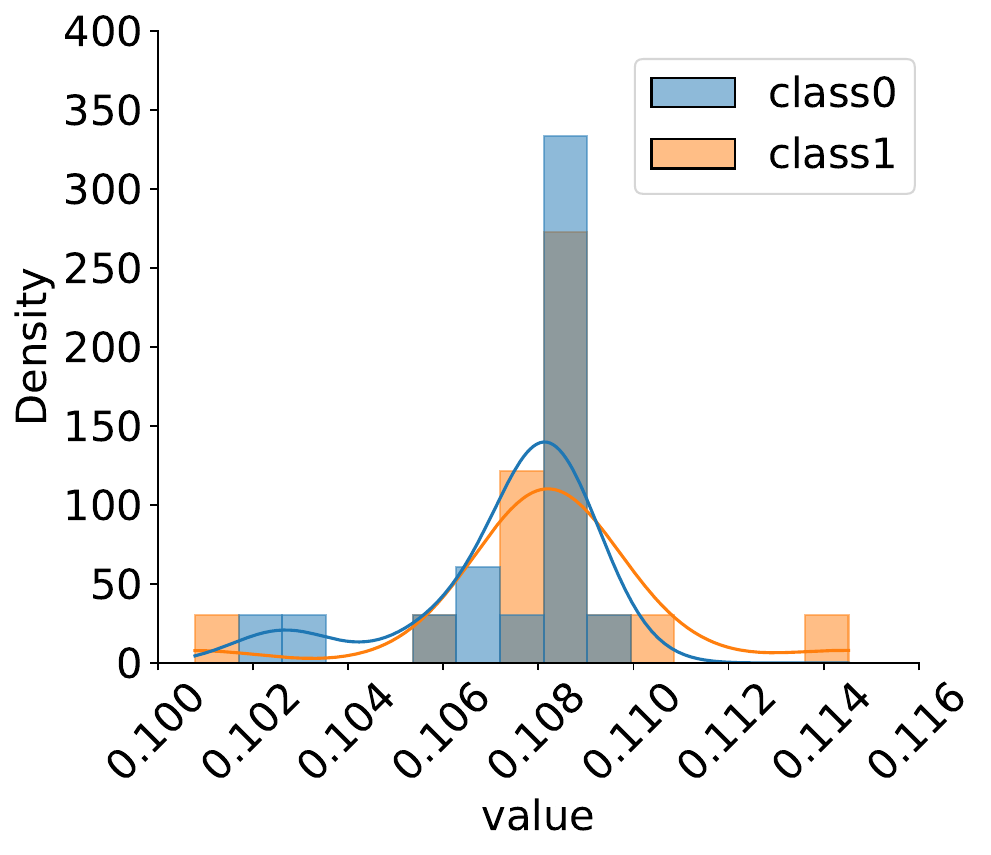}}
\subfigure[\textbf{Last.fm-360K (RAID)}]{\includegraphics[width=4.2cm]{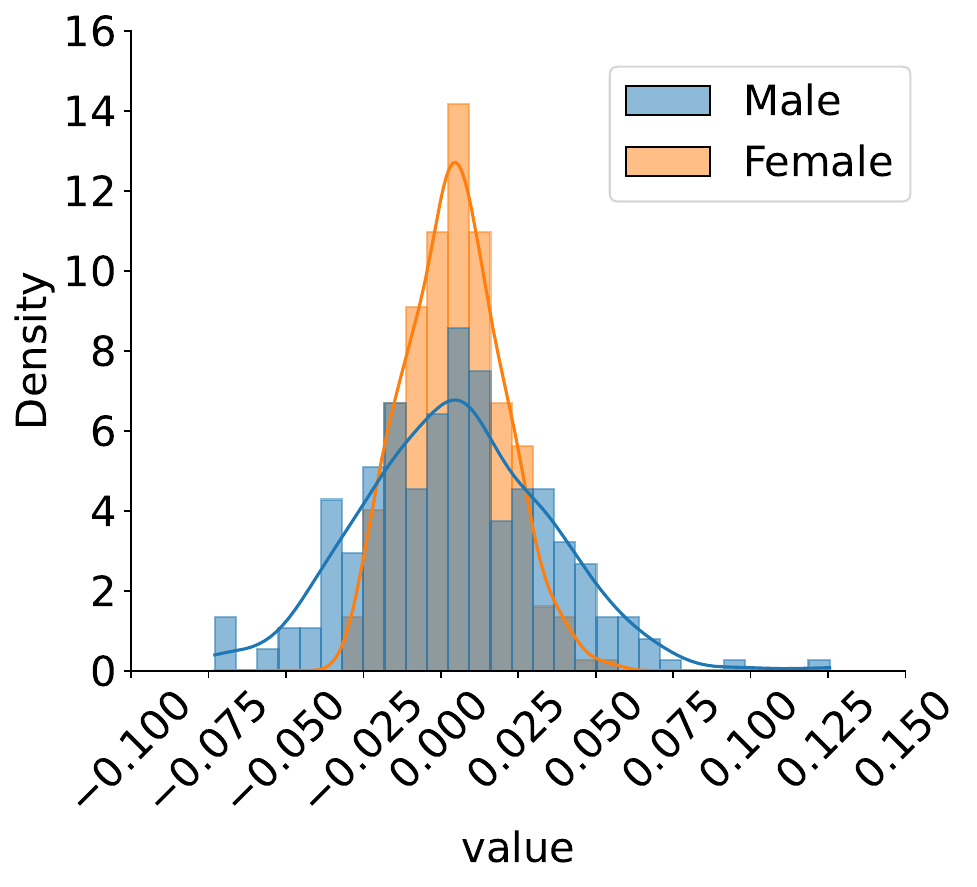}}
\caption{Distribution of user embedding on NCF, where different colors denote different class values (e.g., male vs. female).}
\label{fig:emb_distribution}
\end{figure*}

Existing defense methods lack theoretical guarantees, making it uncertain whether they can effectively safeguard against attacks. 
Our method addresses this limitation. 
We define the defense objective as minimizing the distance between class distributions of protected attributes, thereby preventing attackers from training a classifier for effective separation. 
We validate this on datasets ML-100K, ML-1M, LFM-2B, KuaiSAR, and model NCF by comparing the class distributions of user embeddings in the original model with those after protection, as illustrated in Figure~\ref{fig:emb_distribution}.
We compress the user embeddings to one dimension using PCA, retaining the direction with the maximum variance in the data. 
We present the histogram of class distributions in Figure~\ref{fig:emb_distribution}, with each color representing a different class.
%
%
%
%
As shown in the first row of Figure~\ref{fig:emb_distribution}, there is a noticeable difference between the two class distributions for the original user embedding.
This difference enables the attacker to easily distinguish between the classes, thereby inferring their corresponding attributes.
In contrast, as shown in the second row of Figure~\ref{fig:emb_distribution}, the RAID converges different class distributions towards a single centroid distribution, thereby validating the effectiveness of our defense strategy.

\subsection{Robustness to Adaptive Attack}

In real-life scenarios, numerous models are available for conducting AIA, often rendering the attacker unknown to the defender. 
To better validate the robustness of our method, we also investigate other types of attackers. 
Additionally, we adjust the amount of user data accessible to the attacker. 
Specifically, we use gender and age as target attributes and conduct validations on the ML-1M dataset.

\paragraph{Non-DNN-based Attackers} We explore several machine learning models commonly used in classification tasks as attackers, including Decision Tree (DT), Support Vector Machine (SVM), Naive Bayes (NB), and $k$-Nearest Neighbors (KNN).
We observe from Table~\ref{tab:multi_attack_non_dnn} that
\begin{itemize}[leftmargin=*] \setlength{\itemsep}{-\itemsep}
    \item It is evident that our proposed RAID significantly outperforms other baselines, indicating that our method can more effectively protect attribute information from the recommendation model and safeguard user privacy when facing unknown attacker models. 
    Specifically, DP-PoT, AU-PoT, Adv-InT, and RAID respectively reduced the BAcc by 17.45\%, 36.40\%, 7.31\%, and 40.79\%. 
    In most cases, the BAcc after RAID defense was similar to that of a random attacker.
    \item The defense effectiveness of Adv-InT, which is trained against specific DNN-based inference models, deteriorates when faced with non-DNN-based models. 
    Specifically, when the attacker is an MLP, Adv-InT reduced the BAcc by 9.25\%, but when the attacker was not an MLP, the average reduction was only 7.31\%.
    \item DNN-based attackers (i.e., MLP) generally perform better than other attackers in most scenarios due to their ability to learn nonlinear associations between user embeddings and target attribute labels.
\end{itemize}

\begin{table} [t]
\caption{Performance of attribute inference attack w.r.t. different types of non-DNN-based attacker. The top results are highlighted in \textbf{bold}.
}
\label{tab:multi_attack_non_dnn}
\centering
\resizebox{\linewidth}{!}{
\begin{tabular}{cccccccccc}
\toprule
\multirow{2}{*}{Attribute} & \multirow{2}{*}{Method} & \multicolumn{2}{c}{DT} & \multicolumn{2}{c}{SVM} & \multicolumn{2}{c}{NB} & \multicolumn{2}{c}{KNN} \\
\cmidrule{3-10}
& & F1 & BAcc &  F1 & BAcc & F1 & BAcc & F1 & BAcc \\
\midrule

\multirow{5}{*}{Gender} & Original  & 0.6391 & 0.6379 & 0.7500 & 0.7496 & 0.7271 & 0.7306 & 0.7018 & 0.6981 \\
& DP-PoT   & 0.5378 & 0.5437 & 0.6459 & 0.6358 & 0.6451 & 0.6428 & 0.5617 & 0.5679\\
& AU-PoT   & \textbf{0.5043} & \textbf{0.5152} & 0.4503 & 0.4637 & \textbf{0.4737} & 0.4696 & 0.5042 & 0.5006\\
& Adv-InT   & 0.5224 & 0.5234 & 0.6193 & 0.6264 & 0.5303 & 0.5290 & 0.5870 & 0.5977\\
& RAID   & 0.5151 & 0.5197 & \textbf{0.3995} & \textbf{0.4096} & 0.4762 & \textbf{0.4556} & \textbf{0.4042} & \textbf{0.4292}\\
\midrule

\multirow{5}{*}{Age} & Original  & 0.4646 & 0.4646 & 0.6130 & 0.6130 & 0.5535 & 0.5535 & 0.5094 & 0.5093 \\
& DP-PoT   & 0.3733 & 0.3733 & 0.4591 & 0.4591 & 0.4675 & 0.4674 & 0.4018 & 0.4018\\
& AU-PoT   & 0.3533 & 0.3533 & 0.2587 & 0.2587 & 0.2990 & 0.2989 & 0.2923 & 0.2923\\
& Adv-InT   & 0.5388 & 0.5388 & 0.5428 & 0.5428 & 0.5412 & 0.5412 & 0.6943 & 0.6943\\
& RAID   & \textbf{0.3284} & \textbf{0.3284} & \textbf{0.2277} & \textbf{0.2277} & \textbf{0.2923} & \textbf{0.2923} & \textbf{0.2721} & \textbf{0.2721}\\
\bottomrule
\end{tabular}
}
\end{table}

\paragraph{Ensemble Learning-based Attackers} Ensemble learning is a widely used technique to enhance the performance of classification models. 
We explore several well-recognized ensemble learning-based attackers, including Random Forest (RF), AdaBoost, XGBoost, and Gradient Boosting Decision Tree (GBDT).
As shown from Table~\ref{tab:multi_attack_ensemble}, 
%
%
i) Our proposed RAID framework consistently outperforms the compared baselines in terms of defensive performance against various ensemble learning-based attackers. 
This highlights the robustness of our RAID framework in safeguarding user privacy across different sophisticated attacking methods;
and ii) Comparing the performance of various attackers on the original embedding, the MLP attacker still outperforms those based on ensemble learning. 
This indicates that while ensemble methods are effective, the MLP's ability to capture complex patterns in data makes it a more formidable adversary in scenarios involving AIA.

\begin{table} [t]
\caption{Performance of attribute inference attack w.r.t. different types of ensemble learning-based attackers. The top results are highlighted in \textbf{bold}.
}
\label{tab:multi_attack_ensemble}
\centering
\resizebox{\linewidth}{!}{
\begin{tabular}{cccccccccc}
\toprule
\multirow{2}{*}{Attribute} & \multirow{2}{*}{Method} & \multicolumn{2}{c}{RF} & \multicolumn{2}{c}{AdaBoost} & \multicolumn{2}{c}{XGBoost} & \multicolumn{2}{c}{GBDT} \\
\cmidrule{3-10}
& & F1 & BAcc &  F1 & BAcc & F1 & BAcc & F1 & BAcc \\
\midrule
\multirow{5}{*}{Gender} & Original & 0.6927 & 0.6947 & 0.7373 & 0.7381 & 0.7101 & 0.7124 & 0.7097 & 0.7100 \\
& DP-PoT & 0.5882 & 0.5919 & 0.6321 & 0.6319 & 0.6121 & 0.6115 & 0.6061 & 0.6077 \\
& AU-PoT & 0.5116 & 0.5049 & \textbf{0.5100} & \textbf{0.5009} & 0.5372 & 0.5497 & 0.5314 & 0.5375 \\
& Adv-InT & 0.5214 & 0.5253 & 0.5432 & 0.5319 & 0.5623 & 0.5708 & 0.5555 & 0.5614 \\
& RAID & \textbf{0.4744} & \textbf{0.4947} & 0.5194 & 0.5166 & \textbf{0.5319} & \textbf{0.5348} & \textbf{0.5265} & \textbf{0.5299} \\
\midrule

\multirow{5}{*}{Age} & Original & 0.5111 & 0.5112 & 0.5898 & 0.5898 & 0.5951 & 0.5951 & 0.5863 & 0.5863 \\
& DP-PoT & 0.3983 & 0.3984 & 0.4423 & 0.4423 & 0.4325 & 0.4325 & 0.4201 & 0.4201 \\
& AU-PoT & 0.3481 & 0.3482 & 0.4423 & 0.4423 & 0.4325 & 0.4325 & 0.4201 & 0.4201 \\
& Adv-InT & 0.4509 & 0.4509 & 0.4825 & 0.4825 & 0.5527 & 0.5527 & 0.5558 & 0.5558 \\
& RAID & \textbf{0.3236} & \textbf{0.3236} & \textbf{0.3316} & \textbf{0.3317} & \textbf{0.3440} & \textbf{0.2429} & \textbf{0.3332} & \textbf{0.3333} \\
\bottomrule
\end{tabular}
}
\end{table}

\paragraph{Different layers of MLP} In previous experiments, we employed a two-layer MLP as the attacker and found that the MLP's attack effectiveness was substantially stronger than other attack models. 
To further explore the robustness of our method, in this experiment, we investigate the impact of different MLP architectures. 
Specifically, we examine the effects of varying the number of layers: 1, 2, 3, and 4. 
From the results presented in Table~\ref{tab:multi_attack_mlp_layer}, we observe that the two-layer MLP performs the best among all the compared attackers. 
Additionally, increasing the number of layers does not enhance attack performance. 
We also note that our proposed RAID outperforms other comparative methods in most cases, confirming the effectiveness of RAID in defending against increasingly complex MLP architectures. 
This suggests that our defense mechanism is not only effective against sophisticated attacks but also adaptable to variations in attack model complexities.

\begin{table}[t]
\caption{Performance of attribute inference attack w.r.t. different types of MLP-based attacker. The top results are highlighted in \textbf{bold}. The dimensions of layers are $\{d_{out}\}$, $\{100, d_{out}\}$, $\{100, 64, d_{out}\}$, $\{100,64,32,d_{out}\}$, where $d_{out}$ denotes the count of attribute categories.
}
\label{tab:multi_attack_mlp_layer}
\centering
\resizebox{\linewidth}{!}{
\begin{tabular}{cccccccccc}
\toprule
\multirow{2}{*}{Attribute} & \multirow{2}{*}{Method} & \multicolumn{2}{c}{Layer=1} & \multicolumn{2}{c}{Layer=2} & \multicolumn{2}{c}{Layer=3} & \multicolumn{2}{c}{Layer=4} \\
\cmidrule{3-10}
& & F1 & BAcc &  F1 & BAcc & F1 & BAcc & F1 & BAcc \\
\midrule
\multirow{5}{*}{Gender} & Original  & 0.7481 & 0.7494 & 0.7530 & 0.7541 & 0.7408 & 0.7452 & 0.7451 & 0.7377\\
& DP-PoT   & 0.6469 & 0.6499 & 0.6425 & 0.6419 & 0.6459 & 0.6433 & 0.6090 & 0.6226\\
& AU-PoT   & 0.4683 & 0.4436 & 0.5177 & 0.4980 & 0.5339 & 0.5153 & \textbf{0.3009} & \textbf{0.4962}\\
& Adv-InT   & 0.5258 & 0.5169 & 0.6155 & 0.6101 & 0.6345 & 0.6422 & 0.6349 & 0.6338\\
& RAID   & \textbf{0.4588} & 0.4549 & \textbf{0.4832} & \textbf{0.4709} & \textbf{0.3110} & \textbf{0.5000} & 0.4000 & 0.5000\\
\midrule

\multirow{5}{*}{Age} & Original  & 0.6191 & 0.6191 & 0.6226 & 0.6226 & 0.6147 & 0.6147 & 0.6120 & 0.6120\\
& DP-PoT   & 0.4261 & 0.4261 & 0.4491 & 0.4490 & 0.4419 & 0.4420 & 0.4251 & 0.4251\\
& AU-PoT   & 0.3267 & 0.3267 & 0.3334 & 0.3333 & 0.3331 & 0.3331 & 0.3329 & 0.3329\\
& Adv-InT   & 0.5519 & 0.5519 & 0.7116 & 0.7116 & 0.6558 & 0.6557 & 0.6283 & 0.6283\\
& RAID   & \textbf{0.2990} & \textbf{0.2990} & \textbf{0.3314} & \textbf{0.3341} & \textbf{0.3211} & \textbf{0.3212} & \textbf{0.3298} & \textbf{0.3240}\\
\bottomrule
\end{tabular}
}
\end{table}

\paragraph{Different ratio of leaked data} In real-world scenarios, the amount of data accessible to attackers is often unclear. 
Based on this consideration, we adjust the proportion of user data that the attacker could obtain, specifically setting it to 20\%, 40\%, 60\%, and 80\% for testing on an MLP. 
The experimental results, as shown in Table~\ref{tab:multi_attack_leak_data}, indicate the following observations: 
i) Once the attacker possesses 40\% of the data, they already have the capability to effectively infer user privacy attributes; and
ii) Our method is minimally affected by variations in the percentage of data available to the attacker, showing insensitivity to these changes and consistently achieving effective defensive outcomes. 
This resilience highlights the robustness of our approach to safeguarding user privacy against varying levels of data exposure to attackers.

\begin{figure*}[t]
\centering
\includegraphics[width=13cm]{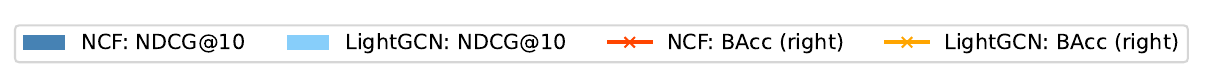}\\
\subfigure[\textbf{ML-1M (Gender)}]{\includegraphics[width=4.2cm]{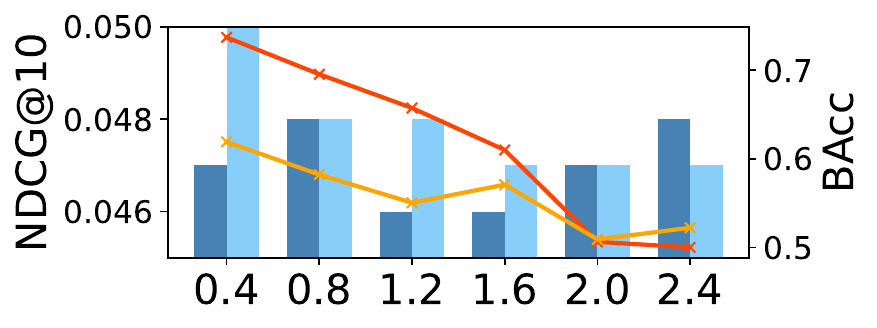} \label{fig:ml-1m_alpha}}
\subfigure[\textbf{LFM-2B (Gender)}]{\includegraphics[width=4.2cm]{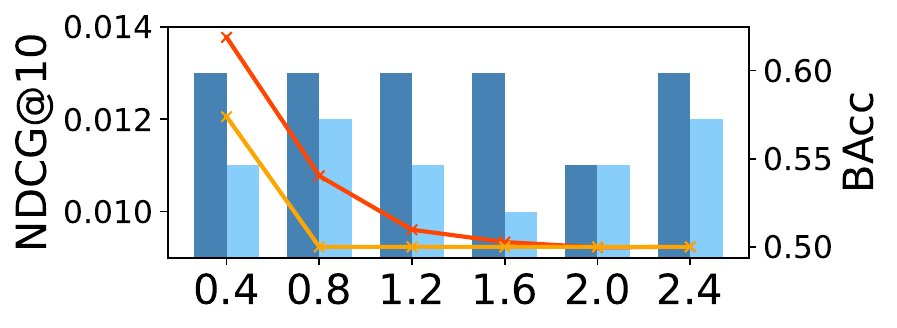} \label{fig:lfm-2b_alpha}}
\subfigure[\textbf{KuaiSAR (Feat1)}]{\includegraphics[width=4.2cm]{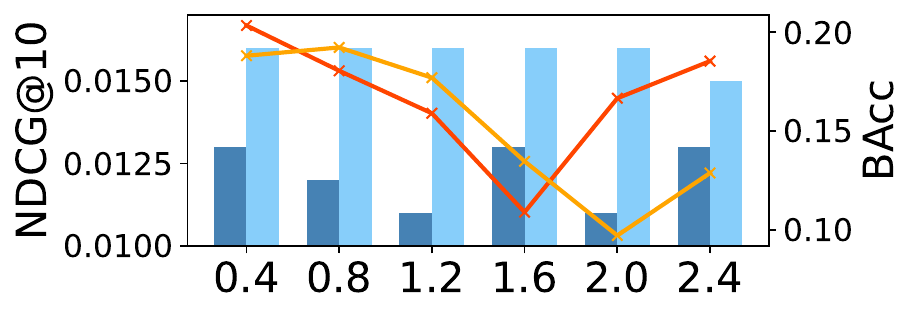} \label{fig:kuaisar_alpha}}
\subfigure[\textbf{Last.fm-360K (Gender)}]{\includegraphics[width=4.2cm]{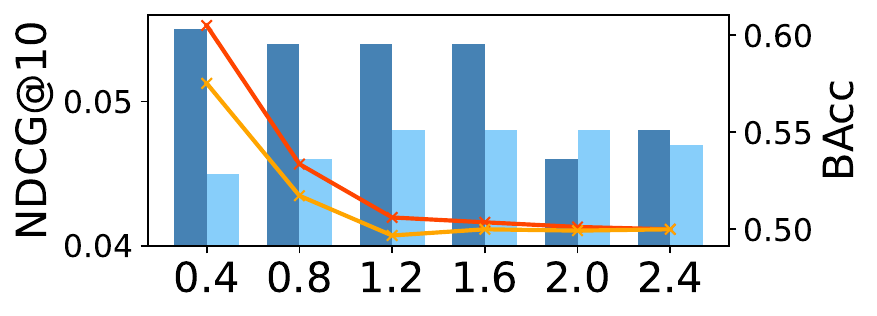} \label{fig:lastfm_alpha}}
\caption{Effect of the hyper-parameter penalization coefficient $\eta$.}
\label{fig:para_eta}
\end{figure*}

\begin{table*}[t]
\caption{Effect of the hyper-parameter frequency $\xi$.
}
\label{tab:para_xi}
\centering
\resizebox{\linewidth}{!}{
\begin{tabular}{cccccccccccccccccc}
\toprule
\multirow{2}{*}{Dataset} & \multirow{2}{*}{Model} & \multicolumn{4}{c}{1} & \multicolumn{4}{c}{2} & \multicolumn{4}{c}{3} & \multicolumn{4}{c}{4} \\
\cmidrule(lr){3-6} \cmidrule(lr){7-10} \cmidrule(lr){11-14} \cmidrule(lr){15-18}
& & F1 & BAcc & HR@10 & NDCG@10 &  F1 & BAcc & HR@10 & NDCG@10 & F1 & BAcc & HR@10 & NDCG@10 & F1 & BAcc & HR@10 & NDCG@10 \\
\midrule

\multirow{3}{*}{ML-1M} & DMF & 0.3110 & 0.4984 & 0.0980 & 0.0420 & 0.3598 & 0.5055 & 0.0830 & 0.0430 & 0.3135 & 0.4972 & 0.0840 & 0.0430 & 0.3454 & 0.5081 & 0.0870 & 0.0420 \\
& NCF & 0.3027 & 0.5045 & 0.0980 & 0.0480 & 0.4444 & 0.5000 & 0.0960 & 0.0490 & 0.3558 & 0.5005 & 0.0960 & 0.0480 & 0.2666 & 0.5000 & 0.0970 & 0.0480 \\
& LightGCN & 0.5340 & 0.5815 & 0.0960 & 0.0480 & 0.6401 & 0.6029 & 0.0960 & 0.0470 & 0.6527 & 0.5979 & 0.0940 & 0.0470 & 0.4593 & 0.5827 & 0.0920 & 0.0460 \\
\midrule

\multirow{3}{*}{LFM-2B} & DMF & 0.4056 & 0.5110 & 0.0170 & 0.0090 & 0.2925 & 0.5005 & 0.0180 & 0.0080 & 0.3884 & 0.5067 & 0.0170 & 0.0080 & 0.3807 & 0.5109 & 0.0160 & 0.0070 \\
& NCF & 0.2121 & 0.5037 & 0.0260 & 0.0130 & 0.3554 & 0.5000 & 0.0240 & 0.0110 & 0.3555 & 0.5000 & 0.0240 & 0.0120 & 0.2769 & 0.5101 & 0.0240 & 0.0110 \\
& LightGCN & 0.3110 & 0.5000 & 0.0240 & 0.0110 & 0.3109 & 0.5000 & 0.0240 & 0.0110 & 0.3973 & 0.4972 & 0.0230 & 0.0110 & 0.3352 & 0.5017 & 0.0230 & 0.0110 \\
\midrule

\multirow{3}{*}{KuaiSAR} & DMF & 0.2175 & 0.2409 & 0.0320 & 0.0160 & 0.2161 & 0.2361 & 0.0310 & 0.0150 & 0.1800 & 0.1958 & 0.0300 & 0.0160 & 0.2543 & 0.2708 & 0.0300 & 0.0150 \\
& NCF & 0.1736 & 0.1861 & 0.0250 & 0.0120 & 0.0972 & 0.1166 & 0.0290 & 0.0140 & 0.1205 & 0.1409 & 0.0250 & 0.0120 & 0.1183 & 0.1333 & 0.0290 & 0.0140 \\
& LightGCN & 0.1018 & 0.1222 & 0.0330 & 0.0160 & 0.1130 & 0.1361 & 0.0330 & 0.0150 & 0.1041 & 0.1250 & 0.0350 & 0.0170 & 0.1018 & 0.1222 & 0.0330 & 0.0160 \\
\midrule

\multirow{3}{*}{Last.fm-360K} & DMF & 0.4918 & 0.5074 & 0.0880 & 0.0490 & 0.2612 & 0.5103 & 0.0730 & 0.0340 & 0.2663 & 0.5000 & 0.0730 & 0.0410 & 0.4672 & 0.5054 & 0.0690 & 0.0430 \\
& NCF & 0.1775 & 0.5152 & 0.0900 & 0.0570 & 0.2222 & 0.5000 & 0.0810 & 0.0520 & 0.1772 & 0.5000 & 0.0950 & 0.0610 & 0.4444 & 0.5000 & 0.0750 & 0.0500 \\
& LightGCN & 0.3994 & 0.5000 & 0.0810 & 0.0480 & 0.3563 & 0.5000 & 0.0810 & 0.0460 & 0.2216 & 0.5000 & 0.0810 & 0.0470 & 0.3105 & 0.5005 & 0.0800 & 0.0450 \\
\bottomrule
\end{tabular}
}
\end{table*}

\begin{table} [t]
\caption{Performance of attribute inference attack w.r.t. different leak ratio. The top results are highlighted in \textbf{bold}.
}
\label{tab:multi_attack_leak_data}
\centering
\resizebox{\linewidth}{!}{
\begin{tabular}{cccccccccc}
\toprule
\multirow{2}{*}{Attribute} & \multirow{2}{*}{Method} & \multicolumn{2}{c}{20\%} & \multicolumn{2}{c}{40\%} & \multicolumn{2}{c}{60\%} & \multicolumn{2}{c}{80\%} \\
\cmidrule{3-10}
& & F1 & BAcc &  F1 & BAcc & F1 & BAcc & F1 & BAcc \\
\midrule
\multirow{5}{*}{Gender} & Original  & 0.6822 & 0.7092 & 0.7243 & 0.7296 & 0.7545 & 0.7524 & 0.7629 & 0.7568\\
& DP-PoT   & 0.7439 & 0.7433 & 0.7530 & 0.7390 & 0.7428 & 0.7470 & 0.7450 & 0.7496\\
& AU-PoT   & 0.4663 & \textbf{0.4650} & 0.5217 & \textbf{0.5014} & 0.5331 & \textbf{0.5108} & 0.5386 & 0.5336\\
& Adv-InT   & 0.5822 & 0.5887 & 0.5971 & 0.5962 & 0.6188 & 0.6037 & 0.6404 & 0.6238\\
& RAID   & \textbf{0.2755} & 0.5000 & \textbf{0.3563} & 0.5146 & \textbf{0.4300} & 0.5250 & \textbf{0.4740} & \textbf{0.5325}\\
\midrule

\multirow{5}{*}{Age} & Original  & 0.5613 & 0.5611 & 0.5891 & 0.5881 & 0.6074 & 0.6045 & 0.6457 & 0.6341\\
& DP-PoT   & 0.5869 & 0.5898 & 0.5947 & 0.5945 & 0.6207 & 0.6200 & 0.6285 & 0.6289\\
& AU-PoT   & \textbf{0.2983} & \textbf{0.2984} & \textbf{0.3032} & \textbf{0.3027} & \textbf{0.3178} & \textbf{0.3177} & \textbf{0.3184} & \textbf{0.3180}\\
& Adv-InT   & 0.4349 & 0.4191 & 0.4632 & 0.4583 & 0.4963 & 0.4955 & 0.5094 & 0.5095\\
& RAID   & 0.3342 & 0.3333 & 0.3321 & 0.3330 & 0.3457 & 0.3451 & 0.3532 & 0.3517\\
\bottomrule
\end{tabular}
}
\end{table}

\subsection{Parameter Sensitivity Analysis}

We conduct an in-depth analysis of the performance fluctuations of our method under different hyper-parameter settings, i.e., the penalization coefficient $\eta$ and the frequency $\xi$ of centroid distribution computation. 
Specifically, we adjust the values of $\eta$ and $\xi$ while keeping other hyper-parameters constant. 

\paragraph{Penalization coefficient $\eta$} Regarding the penalization coefficient $\eta$, as depicted in Figure~\ref{fig:para_eta}, we set $\eta$ values to 0.4, 0.8, 1.2, 1.6, 2.0, and 2.4. 
To measure defensive and recommendation performance, we used BAcc and NDCG@10 as evaluation metrics.
We observe that our proposed RAID exhibits good robustness in terms of NDCG@10 across different $\eta$ values. 
Simultaneously, we find that increasing $\eta$ leads to a decrease in BAcc. 
This observation suggests that RAID can enhance defensive effectiveness with only a slight sacrifice in recommendation performance. 
Additionally, in multi-class scenarios (Figure~\ref{fig:kuaisar_alpha}, 7 classes for Feat1), RAID incurs less performance loss while enhancing defense effectiveness, which might be due to the centroid distribution calculated in multi-class scenarios retaining more information other than the protected attribute compared to the original class-conditional distribution.

\paragraph{Frequency $\xi$} Concerning the frequency $\xi$ of centroid distribution computation, we compare model performance with $\xi$ values of 1, 2, 3, and 4. 
The experimental results, as shown in Table~\ref{tab:para_xi}, demonstrate that our proposed RAID's defensive (i.e., F1 and BAcc) and recommendation (i.e., HR@10 and NDCG@10) performance also exhibit good robustness to different $\xi$ values. 
Although changes were not substantial, the BAcc metric slightly decreased as $\xi$ was reduced. 
Meanwhile, as $\xi$ increased, the training convergence time of the model noticeably decreased.

%% file: section/7-Conclusion.tex
\section{Conclusion}

In this work, we propose a practical two-stage framework named RAID for defending against AIA in recommender systems. 
Specifically, \model{} aligns the attribute distributions to an ideal centroid distribution with constraints to make attributes indistinguishable, thereby defending against AIA while preserving recommendation performance.
In phase I, \model{} is trained directly on the training data to achieve initial recommendation performance. In phase II, \model{} first obtains the ideal centroid distribution by solving a constrained Wasserstein-2 centroid optimization problem based on the distribution of each class in the protected attribute. 
Subsequently, during the training process, \model{} employs optimal transport to align the distribution of each class toward this centroid distribution, thereby defending against AIA.
Our empirical results on real-world datasets and mainstream recommendation models demonstrate that: i) Our proposed method significantly outperforms existing baselines in terms of defense effectiveness; ii) Our proposed method has minimal impact on the recommendation performance of the model; iii) Our proposed method shows significant efficiency improvements and easier convergence compared to adversarial training methods; and iv) Our proposed method is applicable to both binary and multi-class attributes.
Notably, \model{} is effective against gray-box attacks and can be extended to address fairness issues in recommender systems.

%
%

%% file: section/Appendix.tex

\section{Proofs and Optimization Method} \label{app:proofs}

\subsection{Proof of Lemma~\ref{lem:com_wb}} We first present some prior results about the optimal transport problem.\\
With probability measures $\mu,\nu$ on $\mathcal{X}$, we consider the the Wasserstein distance $\mathcal{W}_2^2(\mu,\nu)$, which is defined by
\begin{equation}
    \mathcal{W}_2^2(\mu, \nu) {:=} \min_{\pi \in \Pi(\mu, \nu)} \int_{\mathcal{X}^2} \|x - y\|^2_2 \, dT(x, y),
    \label{W2-distance}
\end{equation}
where $\Pi(\mu, \nu)$ is the set of transport plans with marginals $\mu$ and $\nu$ on $\mathcal{X}^2$.
Note that this expression is equivalent to Eq.~\eqref{eq:kanto} with the substitution of $c(x,y)$ by $\| x-y \|^2_2$.
Previous work~\cite{otdual-genevay2016stochastic} shows that dual and semi-dual formulation of wasserstein distance Eq.~\eqref{W2-distance} are defined by
\begin{align*}
    \mathcal{W}_2^2(\mu, \nu) &= \max_{\substack{f, g \in \mathcal{C}(\mathcal{X}) \\ f \oplus g \leq c}} \int_X f(x) d\mu(x) + \int_X g(y) d\nu(y), \tag{dual}\\
    &= \max_{g \in \mathcal{C}(\mathcal{X})}  \int_X g^c(x) d\mu(x) + \int_X g(y) d\nu(y),\tag{semi-dual}
\end{align*}
where $c(x,y) = \| x-y \|^2_2$, $\mathcal{C}(\mathcal{X})$ denotes continuous functions on $\mathcal{X}$, and $g^c(x) := \min_{y \in \mathcal{X}} \{c(x,y)-g(y)\}$ is the $c$-transform of $g$.
Given empirical probability measures (histograms)
\[
\mu = \sum_{n=1}^{N_{\mu}} \alpha_n \delta_{y_n^\mu}, \sum_{n=1}^{N_{\mu}} \alpha_n =1,\quad \nu = \sum_{n=1}^{N_{\nu}} \beta_n \delta_{y_n^\nu},\sum_{n=1}^{N_{\nu}} \beta_n =1,
\]
the discretized semi-dual problem~\cite{fast-wb-cuturi2014fast} is
\begin{align}
 \max_{g \in \mathbb{R}^{N_{\mu}}}  \sum_{n=1}^{N_{\mu}} g^c(x_n) \alpha_n + \sum_{n=1}^{N_{\nu}} g(y_n) \beta_n,
 \label{discrete-semidual}
\end{align}
where $g^c(x_n):=\min_{q\in [N_{\nu}]} \{ c(x_n, y_q) - g(y_q )\}$.
Firstly, we derive that the Wasserstein barycenter defined in Eq.~\eqref{eq:wb} has the dual formulation as presented in Eq.~\eqref{dual-problem} when we consider its empirical estimation.
In practical, the $\hat{\mathcal{P}}(Y^1), \ldots, \hat{\mathcal{P}}(Y^K)$ are given in the form of $K$ sample-based empirical probability measures supported on a compact subset $\mathcal{X} \subset \mathbb{R}^d$ with $d \geq 1$,
\begin{align}
\label{emp-class}
\hat{\mathcal{P}}(Y^i) = \frac{1}{N_i}\sum_{n=1}^{N_i} \delta_{y_n^i}.
\end{align}
For multi-dimensional barycenter estimation, we follow prior research~\cite{fast-wb-cuturi2014fast,conti-wb-jayaraman2022attribute,2-wb-korotin2021continuous} to employ fixed-point iterations for computing empirical Wasserstein barycenter in given fix support $\{ \bar{y}_q \}_{q=1}^{\bar{N}}$ as an approximation to the unknown true Wasserstein barycenter distribution.
The empirical Wasserstein barycenter can be estimated by 
\begin{equation}
\label{emp-wb}
\hat{\mathcal{P}}^* =  \sum_{n=1}^{\hat{N}} \alpha_n \delta_{\bar{y}_n},\quad \sum_{n=1}^{\hat{N}} \alpha_n =1, \alpha_n \geq 0.
\end{equation}
Denoting the sample-based distributions $\hat{\mathcal{P}}(Y^i)$ by $\hat{\mathcal{P}}^i$, with $\mathcal{M}_1^+(\mathcal{X})$ presenting the set of all probability measures on $\mathcal{X}$, the semi-dual problem of penalized Wasserstein barycenter problem can be formulated as 
\begin{equation*}
\label{continuouos-semi-dual}
\begin{split}
&\min_{\hat{\mathcal{P}}^* \in \mathcal{M}_1^+(\mathcal{X})} \ \max_{g_1,\cdots,g_K} H(\hat{\mathcal{P}}^*,\mathbf{g})\\
&\qquad \quad =\sum_{i=1}^K \lambda_i \int_\mathcal{X} g_i^c(y)d\hat{\mathcal{P}}^i (y) + \int_\mathcal{X} \sum_{i=1}^K \lambda_i g_i(y)d\hat{\mathcal{P}}^* (y)\\
&\qquad \quad\quad + \tau\int_\mathcal{X} \log(\frac{d\hat{\mathcal{P}}^*}{dy})d\hat{\mathcal{P}}^* (y). \\
\end{split}
\end{equation*}
Plugging in the empirical distributions $\hat{\mathcal{P}}^*, \hat{\mathcal{P}}^i,i\in [K]$ represented by Eq.~\eqref{emp-class} and Eq.~\eqref{emp-wb}, we obtain the empirical penalized Wasserstein barycenter problem 
\begin{equation}
    \label{emp-semi-dual}
    \begin{split}
    &\min_{\substack{ \alpha_1,\cdots,\alpha_{\hat{N}} \geq 0\\ \sum_{n=1}^{\hat{N}} \alpha_n =1} } \ \max_{g_1,\cdots,g_K} \hat{H}(\mathbf{\alpha},\mathbf{g})\\
    &\qquad \quad =\sum_{i=1}^K \lambda_i \frac{1}{N_i} \sum_{n=1}^{{N}_i} g_i^c({y}_n^i) + \sum_{n=1}^{\hat{N}} \sum_{i=1}^K \lambda_i g_i(\bar{y}_n) \alpha_n\\
    &\qquad \quad \quad + \tau \sum_{n=1}^{\hat{N}} \alpha_n\log(\alpha_n),\\
    \end{split}
\end{equation}
where $g^{c}_i(y_n^i)$ is defined by $\min_{p=1,\cdots,\hat{N}} \{ \| y_n^i - \bar{y}_p \|^2_2 - g_i(\bar{y}_p)\}$. 
%
%
It is explicit that Eq.~\eqref{emp-semi-dual} is concave to $\alpha$ and convex to $\textbf{g}$. Since the domain of $\alpha$ is convex and compact, and the domains of $g_i$'s are convex,  we can switch min and max according to Sion’s minimax theorem,
\begin{equation}
    \label{switch-emp-semi-dual}
    \begin{split}
    & \max_{g_1,\cdots,g_K} \ \min_{\substack{ \alpha_1,\cdots,\alpha_{\hat{N}} \geq 0\\ \sum_{n=1}^{\hat{N}} \alpha_n =1} } \hat{H}(\mathbf{\alpha},\mathbf{g}).\\
    \end{split}
\end{equation}
The next step is to solve the inner minimum problem with respect to $\alpha$ in Eq.~\eqref{switch-emp-semi-dual}. Introducing Lagrangian multiplier $\kappa$ and $\rho=(\rho_1,\cdots,\rho_{\hat{N}})$,  we investigate the Lagrangian form,
\begin{equation}
\label{Lagrangian}
\begin{split}
 & L(\alpha,\kappa,\rho)\\
 & \ = \sum_{n=1}^{\hat{N}}\left( \sum_{i=1}^K \lambda_i g_i(\bar{y}_n) + \tau \log(\alpha_n) \right)\alpha_n + \kappa\left(\sum_{n=1}^{\hat{N}}\alpha_n-1\right) \\
 & \qquad - \sum_{n=1}^{\hat{N}}\rho_n \alpha_n.\\
\end{split}
\end{equation}
By Karush-Kuhn-Tucker (KKT) condition, for optimal solution $(\alpha^*,\kappa^*,\rho^*)$, we have
\begin{align*}
&\frac{\partial L}{\partial \alpha^*_n} = \sum_{i=1}^K \lambda_i g_i(\bar{y}_n) + \tau \left(\log(\alpha^*_n) + 1 \right) +\kappa^* - \rho^*_n=0,\\
&\rho_n^* \alpha_n^*=0,\ \rho_n^* \geq 0,\quad n \in [\hat{N}],\\
&\left(\sum_{n=1}^{\hat{N}}\alpha^*_n-1\right) = 0.\\
\end{align*}
Solving these equations, the optimal solution  $\alpha^*$ can be formulated as 
\begin{align}
\label{optimal-alpha}
\alpha_n^* = \frac{\exp\left( -\frac{c_n}{\tau} \right)}{\sum_{q=1}^{\hat{N}} \exp\left( -\frac{c_q}{\tau} \right)},\quad c_n = \sum_{i=1}^K \lambda_i g_i(\bar{y}_n) .
\end{align}
Plugging Eq.~\eqref{optimal-alpha} into Eq.~\eqref{switch-emp-semi-dual}, the empirical penalized Wasserstein barycenter problem turn to 
%
\begin{align*}
    \hat{H}(\mathbf{\alpha^*}(\mathbf{g}),\mathbf{g}) &= \sum_{n=1}^{\hat{N}}\left( \sum_{i=1}^K \lambda_i g_i(\bar{y}_n) + \tau \log(\alpha_n^*) \right)\alpha_n^*\\
    & \quad +\sum_{i=1}^K \lambda_i \frac{1}{N_i} \sum_{n=1}^{{N}_i} g_i^c({y}_n^i)\\
    & = \sum_{n=1}^{\hat{N}}-\tau \log \left( \sum_{n=1}^{\hat{N}} \exp \left(-\frac{1}{\tau} \sum_{i=1}^K \lambda_i g_i(\bar{y}_n)\right) \right)\alpha_n^*\\
    & \quad +\sum_{i=1}^K \lambda_i \frac{1}{N_i} \sum_{n=1}^{{N}_i} g_i^c({y}_n^i)\\
    & = -\tau \log \left( \sum_{n=1}^{\hat{N}} \exp \left(-\frac{1}{\tau} \sum_{i=1}^K \lambda_i g_i(\bar{y}_n)\right) \right)\\
    & \quad +\sum_{i=1}^K \lambda_i \frac{1}{N_i} \sum_{n=1}^{{N}_i} g_i^c({y}_n^i).\\
\end{align*}
\normalsize
Consequently, we derive the dual problem of penalized Wasserstein barycenter as presented in Eq.~\eqref{dual-problem}. After we solve $\max_{g_1,\cdots,g_K}\hat{H}(\mathbf{\alpha}^*(\mathbf{g}),\mathbf{g})$, the empirical barycenter $\hat{\mathcal{P}}^*$ cen be recovered from $\mathbf{g}$ by implementing  Eq.~\eqref{optimal-alpha}. Then the optimal couplings $\{T_i^*\}_{i=1}^K$ are able to be computed.
Next, we prove that $\hat{H}(\mathbf{\alpha}^*(\mathbf{g}),\mathbf{g})$ is concave with respect to $\mathbf{g}=(g_i)_{i=1}^K$. Let $\hat{H}(\mathbf{\alpha}^*(\mathbf{g}),\mathbf{g})=F(\mathbf{g})+G(\mathbf{g})$,
\begin{align*}
& F(\mathbf{g}) =  -\tau \log \left( \sum_{n=1}^{\hat{N}} \exp \left(-\frac{1}{\tau} \sum_{i=1}^K \lambda_i g_i(\bar{y}_n)\right) \right),\\
& G(\mathbf{g}) =\sum_{i=1}^K \lambda_i \frac{1}{N_i} \sum_{n=1}^{{N}_i} g_i^c({y}_n^i).\\
\end{align*}
For $F(\mathbf{g})$, log-sum-exp function is a classical convex function, and inside the log-sum-exp function is a linear combination 
 which monotonically non-increasing in each component. The composition of the log-sum-exp function and the linear function here remains a convex function. Since the convex function multiplied by a negative constant is concave, $F(\mathbf{g})$ is concave function to $\mathbf{g}$. 
 For $G(\mathbf{g})$, note that $g^{c}_i(y_n^i) = \min_{p=1,\cdots,\hat{N}} \{ \| y_n^i - \bar{y}_p \|^2_2 - g_i(\bar{y}_p)\}$, $g^{c}_i(y_n^i)$ is concave to $\mathbf{g}$ due to the fact that the min operation preserves concavity. It is explicit that $G$ is concave, because $G(\mathbf{g})$ is a linear combination of $g^{c}_i(y_n^i)$.
Consequently, $\hat{H}(\mathbf{\alpha}^*(\mathbf{g}),\mathbf{g})=F(\mathbf{g})+G(\mathbf{g})$ is concave with respect to $\mathbf{g}$.
Additionally, the dual problem can be reformulated as a convex optimization problem by considering $ \min_{g_1,\cdots,g_K} -\hat{H}(\mathbf{\alpha}^*(\mathbf{g}),\mathbf{g})$.
\subsection{Numerical Optimization Scheme} Since we have established that the dual form Eq.~\eqref{dual-problem} of the empirical penalized Wasserstein barycenter problem can be reformulated into an unconstrained convex optimization problem. Subgradient descent can be employed to solve for the optimal $\mathbf{g}$.
\\
\\
\subsection{Proof of Lemma~\ref{lem:com_g}} Given empirical distributions $\hat{\mathcal{P}}(Y^i)$ and $\hat{\mathcal{P}}^*$ with fix supports $\{y^{i}_{p}\}_{p=1}^{N^{i}}$ and $\{\bar{y}_q \}_{q=1}^{\bar{N}}$. The optimal couplings ${T^*_i}$ between $\hat{\mathcal{P}}(Y^i)$ and $\hat{\mathcal{P}}^*$ can be calculated with fixed input histograms. In this paper, $\{y^{i}_{p}\}_{p=1}^{N^{i}}$ represents the embedding features of sample $x_n^i$ extracted with model $f(\theta)$. The embedding distributions can be written as
\begin{align}
\label{emp-embedding}
\hat{\mathcal{P}}_{\theta}(Y^i) = \frac{1}{N_i}\sum_{n=1}^{N_i} \delta_{y_n^i(\theta)}, i \in [K].
\end{align}
Hence, given the barycenter defined in the 2-Wasserstein distance problem turn to 
%
\begin{equation}
    \mathcal{W}_2^2(\hat{\mathcal{P}}_{\theta}(Y^i),\hat{\mathcal{P}}^*) = \min_{T_i\in U(\mathbf{1}_{N_{i}}/N_{i},\mathbf{\alpha})} \left\{\sum_{p=1}^{N_i} \sum_{q=1}^{\bar{N}} T_i(p,q) c(y_p^i(\theta),\bar{y}_q ) \right\},
    \label{eq:theta-w2}
\end{equation}
\normalsize
where $c(y_p^i(\theta),\bar{y}_q) =\| y_p^i(\theta)-\bar{y}_q \|^2$ and $U(\mathbf{1}_{N_{i}}/N_{i},\mathbf{\alpha})$ denotes that $\{T(p,q): \sum_{p=1}^{N_i}T(p,q)=\alpha_q, \ \sum_{q=1}^{\bar{N}}T(p,q)=1/N_i \}$.
By envelope theorem~\cite{envelope}, 
\begin{equation}
    \nabla_{\theta} \mathcal{W}_2^2(\hat{\mathcal{P}}_{\theta}(Y^i),\hat{\mathcal{P}}^*) = \sum_{p=1}^{N_i} \sum_{q=1}^{\bar{N}} T_i^*(p,q) \nabla_{\theta} c(y_p^i(\theta),\bar{y}_q ).
    \label{eq:theta-w3}
\end{equation}
Here, $T^*$ is the transport plan that minimizes the objective in Equation~\eqref{eq:theta-w2}. Similar techniques have been adopted in existing work (Lemma 1 of~\cite{fairgeneral}, Theorem 3 of~\cite{wassersteinGAN}) to derive the gradient of the Wasserstein distance.
Given 2-Wasserstein distance, using the chain rule, the gradient in~\eqref{eq:theta-w3} can be formulated as 
\begin{align}
    &\nabla_{\theta} \mathcal{W}_2^2(\hat{\mathcal{P}}_{\theta}(Y^i),\hat{\mathcal{P}}^*)\\
    &= \sum_{p=1}^{N_i} 2{\left[ \partial y_p^i(\theta) \right]}^{\top} \left( \sum_{q=1}^{\bar{N}} T_{i}^{*}(p,q) (y_p^i(\theta) - \bar{y}_q) \right)_{p=1}^{N^i}\\
    &=\sum_{p=1}^{N_i} 2{\left[ \partial y_p^i(\theta) \right]}^{\top} \left( \frac{y_p^i}{N^i} - \sum_{q=1}^{\bar{N}} T_{i}^{*}(p,q) \bar{y}_q \right)_{p=1}^{N^i}.
    \label{eq:theta-w4}
\end{align}
Plugging the gradient of 2-Wasserstein distance into~\eqref{eq:empirical}, we get the representation in Lemma~\ref{lem:com_g}.

\section{Experimental Settings}\label{sec:exp_details}

\subsection{Dataset Pre-processing}
For all used datasets, we first filter out users without valid attribute information. 
Following the settings described in previous studies~\cite{he2017neural,xue2017deep}, for ML-1M, LFM-2B, and KuaiSAR, we retain only users with at least 5 interactions and items receiving at least 5 interactions from distinct users. For Last.fm-360K, we retain users with at least 45 interactions and items receiving more than 45 interactions from distinct users.
The characteristics of these datasets are summarized in Table~\ref{tab:dataset}. 
To evaluate recommendation performance, we use the widely employed leave-one-out testing~\cite{he2017neural}, where we retain the last two items for each user (sorted by interaction time), using one as a validation item and the other as a test item.
Regarding attribute data, we utilize two attributes from MovieLens, LFM-2B, and Last.fm-360K datasets: age and gender. 
Based on the previous studies~\cite{xue2017deep,moon2023feature}, we categorize the age attribute into three groups: above 45 years, below 35 years, and between 35 to 45 years, while the provided gender attribute is limited to female and male. 
For KuaiSAR, we utilize encrypted one-hot anonymous categories of users as the target attributes.
In each dataset, we randomly sample 90\% of the interaction data to serve as \textit{the training set}, with the remaining 10\% set aside as \textit{the testing set}, which is not used during training. 

\subsection{Parameter Settings} Our detailed experimental setup is as follows:
\begin{itemize}[leftmargin=*] \setlength{\itemsep}{-\itemsep}
    \item \textbf{Recommendation Models}: All model parameters are initialized with a Gaussian distribution $\mathcal{N}(0, {0.01}^2)$.
    To obtain the optimal performance, we use grid search to tune the hyper-parameters. For model-specific hyper-parameters, we follow the suggestions from their original papers. 
    Specifically, we employ the SGD optimizer with a learning rate of $1e-4$ and an embedding size of 32. 
    The number of epochs is set to 10 for DMF, 20 for NCF, and 200 for LightGCN.
    For negative sampling, following prior work, we sample four items per iteration that the user has not interacted with, generating four negative samples for each positive interaction.
    Furthermore, for our RAID framework, we set the penalty coefficient $\eta=1$, with the number of optimization rounds for the first and second phases set at 5 and 15, respectively. 
    The frequency $\xi$ of computing the central distribution is set at 4 and the penalty coefficient $\tau$ for the centroids is set to 1.
    \item \textbf{Attacker}:     For RF, we set the n\_estimators to 100 and the max\_depth to 20. 
    For GBDT, we set the n\_estimators to 100. All these three models are implemented with scikit-learn 1.1.3~\footnote{https://scikit-learn.org/}.
    For XGBoost, we use the xgboost package~\footnote{https://github.com/dmlc/xgboost/tree/stable}, setting the hyper-parameters as their default values. 
    For AdaBoost, we set the n\_estimators to 50. 
    For MLP, we set the dimension of its hidden layer to 100, with a softmax layer serving as its output layer.
    Additionally, we set the L2 regularization weight to 1.0, the initial learning rate to $1e-3$, and the maximal iteration to 500, leaving the other hyper-parameters at their defaults.

\end{itemize}

\section{More Discussions}\label{sec:dis_app}

\subsection{Defend Against Attacks on Multiple Attributes}
In real-world scenarios, attackers may target multiple sensitive attributes simultaneously.
To evaluate RAID's capability to defend against multiple attribute attacks, we compare it with other baselines.
For evaluation, we implement three unlearning strategies on the NCF model with the ML-1M dataset:
i) AU-PoT and RAID employ sequential attribute unlearning, 
ii) Adv-InT (also known as AdvX) implements simultaneous multi-attribute unlearning through adversarial training with multiple discriminators~\cite{escobedo2024simultaneous}.
We then assess both recommendation performance and defense capability against multi-attribute attacks.
The experimental results, as shown in Table~\ref{tab:multiple_attribute}, indicate the following observations: 
i) RAID achieves the lowest F1 and BAcc scores against two-attribute attacks compared to other methods;
ii) Both RAID and AU-PoT maintain recommendation performance (HR/NDCG), with only marginal degradation when sequentially unlearning Age followed by Gender.
These results demonstrate RAID's effectiveness in protecting against multi-attribute attacks while preserving recommendation performance.

\subsection{Limitations and Future Work}

Given the similarities between recommender system models and those used in social networks and mobile devices, our method can be straightforwardly extended to these domains.
However, it is important to note that our assumptions regarding the attacker in recommender systems have certain limitations. 
We assume that the attacker can access the model's parameters or the outcomes of the recommendations, rather than the original training data. 
This assumption may have restrictions in real-world scenarios, and future research could further explore this aspect of the attack model. 
Additionally, our method currently only handles attributes that are discrete variables and cannot manage continuous variable attributes that have not been discretized. 
Addressing more general vector attributes is also a direction worthy of improvement in future research.

\begin{table}[htbp]
  \centering
  \caption{Results of recommendation performance and attribute inference attack on defending against attacks on multiple attributes. The top results are highlighted in bold.}
  \resizebox{0.85\linewidth}{!}{
    \begin{tabular}{cccccc}
    \toprule
    Attribute & Method & HR@10 & NDCG@10 & F1    & BAcc \\
    \midrule
    \multirow{5}[2]{*}{Gender} & Original & 0.1040  & 0.0520  & 0.7501  & 0.7487  \\
          & DP-PoT & 0.0490  & 0.0240  & 0.4425  & \textbf{0.4425 } \\
          & AU-PoT & \textbf{0.1000 } & \textbf{0.0510 } & 0.4868  & 0.4910  \\
          & Adv-InT & 0.0970  & 0.0490  & 0.6180  & 0.6106  \\
          & RAID  & 0.0890  & 0.0460  & \textbf{0.2666 } & 0.5000  \\
    \midrule
    \multirow{5}[2]{*}{Gender, Age} & Original & 0.0820  & 0.0410  & 0.7162  & 0.7188  \\
          & DP-PoT & 0.0220  & 0.0097  & 0.4777  & 0.4776  \\
          & AU-PoT & \textbf{0.0840 } & \textbf{0.0417 } & 0.4895  & 0.4345  \\
          & Adv-InT & 0.0833  & 0.0410  & 0.4962  & 0.4665  \\
          & RAID  & 0.0835  & 0.0410  & \textbf{0.4368 } & \textbf{0.4334 } \\
    \bottomrule
    \end{tabular}
  }
  \label{tab:multiple_attribute}
\end{table}

\section{More Empirical Results}\label{sec:app_res}

\paragraph{Recommendation Performance}To verify the effectiveness of the RAID framework in maintaining recommendation performance, we employ NDCG and HR for evaluation, truncating the ranking lists at 5, 10, 15, and 20. 
We report the results of the ranking lists at 15 and 20 in Table~\ref{tab:rec_performance_15_20}. 

\begin{figure*}
\centering
\subfigure[\textbf{ML-1M}]{\includegraphics[width=3.7cm]{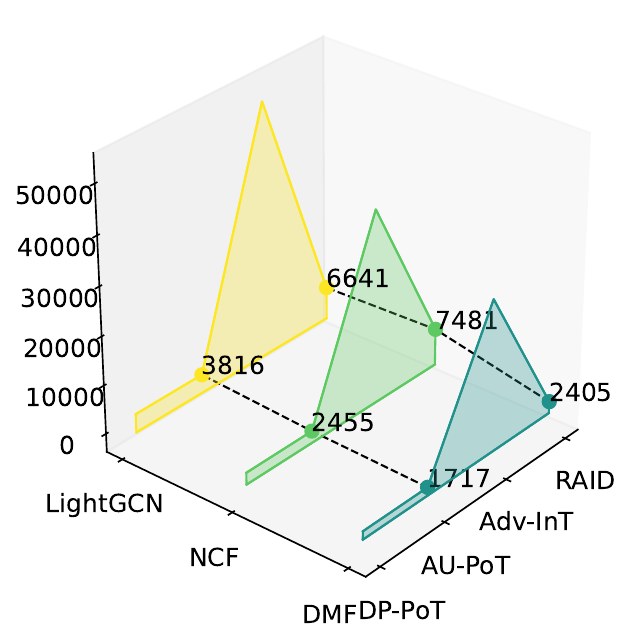}}
\hspace{0.65cm}
\subfigure[\textbf{LFM-2B}]{\includegraphics[width=3.7cm]{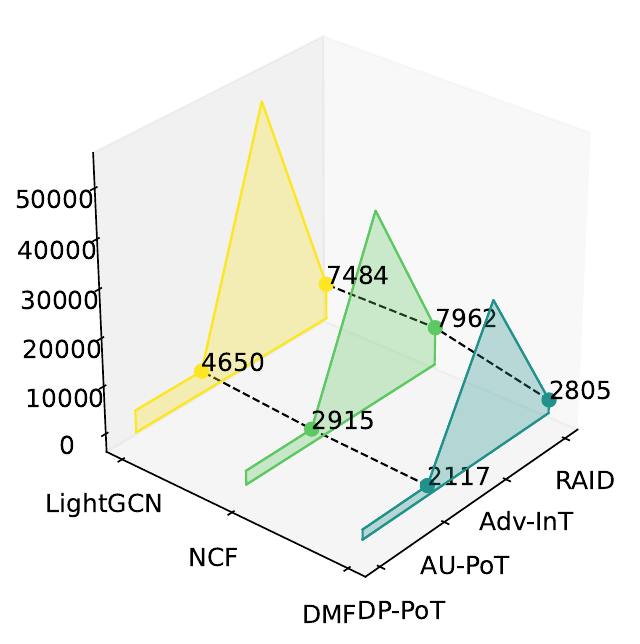}}
\hspace{0.65cm}
\subfigure[\textbf{KuaiSAR}]{\includegraphics[width=3.7cm]{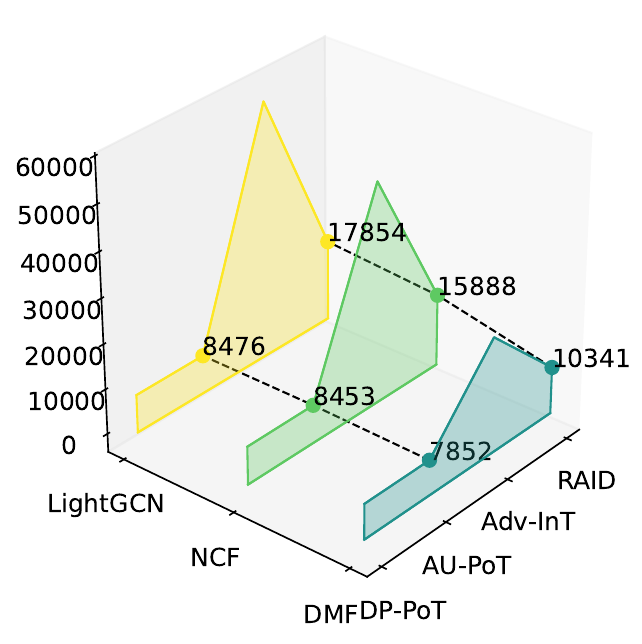}}
\hspace{0.65cm}
\subfigure[\textbf{Last.fm-360K}]{\includegraphics[width=3.7cm]{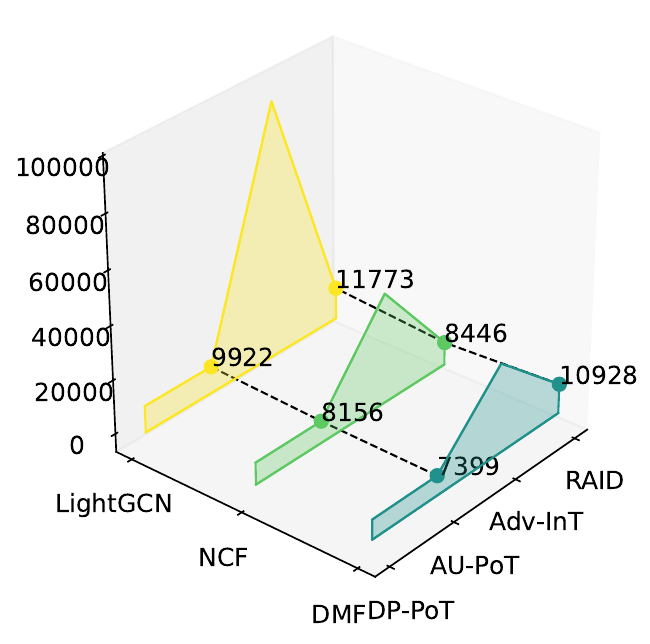}}
\caption{Running time of unlearning methods.}
\label{fig:efficiency}
\end{figure*}

\paragraph{Efficiency} Compared to other InT methods, such as Adv-InT, our method exhibits significant efficiency improvements. 
%
%
Under the conditions of maintaining the same learning rate and setting an early stopping strategy, i.e., terminating training when the sum of defense loss and recommendation loss convergence, we conduct a comparison of all unlearning methods across each dataset and model. 
We report the running times in Figure~\ref{fig:efficiency}. 
From it, we observe that our proposed RAID significantly outperforms Adv-InT in terms of efficiency and stability. 
This is primarily because Adv-InT tends to be unstable during training, prone to fluctuations, and thus difficult to converge. 
In contrast, our method is more stable during parameter updates. 
%
%
%

\begin{table*}[t]
\centering
\caption{Results of recommendation performance with truncating the ranking list for these metrics at 15 and 20. The top results are highlighted in \textbf{bold}.}
\label{tab:rec_performance_15_20}
\resizebox{\linewidth}{!}{
\begin{tabular}{lcccccccccccccc}
\toprule
\multirow{2}{*}{Dataset} & \multirow{2}{*}{Attribute} & \multirow{2}{*}{Method} & \multicolumn{4}{c}{DMF} & \multicolumn{4}{c}{NCF} & \multicolumn{4}{c}{LightGCN} \\ 
\cmidrule(lr){4-7} \cmidrule(lr){8-11} \cmidrule(lr){12-15}
& & & HR@15 & NDCG@15 & HR@20 & NDCG@20 & HR@15 & NDCG@15 & HR@20 & NDCG@20 & HR@15 & NDCG@15 & HR@20 & NDCG@20 \\
\midrule


\multirow{10}{*}{ML-1M} & \multirow{5}{*}{Gender} 
& Original & 0.1390 & 0.0620 & 0.1670 & 0.0680 & 0.1430 & 0.0640 & 0.1730 & 0.0710 & 0.1310 & 0.0580 & 0.1590 & 0.0580 \\
&  & DP-PoT & 0.0630 & 0.0270 & 0.0790 & 0.0310 & 0.0240 & 0.0090 & 0.0310 & 0.0110 & 0.0560 & 0.0260 & 0.0640 & 0.0270 \\
&  & AU-PoT & \textbf{0.1340} & \textbf{0.0590} & \textbf{0.1630} & \textbf{0.0660} & 0.1080 & 0.0490 & 0.1350 & 0.0560 & 0.1040 & 0.0480 & 0.1280 & 0.0540 \\
&  & Adv-InT & 0.1260 & 0.0560 & 0.1550 & 0.0630 & 0.1300 & \textbf{0.0560} & \textbf{0.1620} & \textbf{0.0630} & 0.1120 & 0.0510 & 0.1390 & 0.0570 \\
&  & RAID & 0.1230 & 0.0540 & 0.1470 & 0.0600 & \textbf{0.1310} & \textbf{0.0560} & 0.1600 & \textbf{0.0630} & \textbf{0.1160} & \textbf{0.0520} & \textbf{0.1450} & \textbf{0.0580} \\ \cmidrule{2-15}
& \multirow{5}{*}{Age} 
& Original & 0.1390 & 0.0620 & 0.1670 & 0.0680 & 0.1430 & 0.0640 & 0.1730 & 0.0710 & 0.1310 & 0.0580 & 0.1590 & 0.0580 \\
&  & DP-PoT & 0.0600 & 0.0240 & 0.0750 & 0.0280 & 0.0240 & 0.0090 & 0.0310 & 0.0110 & 0.0560 & 0.0250 & 0.0650 & 0.0270 \\
&  & AU-PoT & 0.1150 & 0.0510 & 0.1470 & 0.0590 & 0.1250 & 0.0570 & 0.1550 & 0.0640 & 0.1040 & 0.0480 & 0.1270 & 0.0530 \\
&  & Adv-InT & 0.1140 & 0.0510 & 0.1380 & 0.0570 & 0.1320 & 0.0560 & 0.1600 & 0.0630 & 0.0980 & 0.0450 & 0.1170 & 0.0490 \\
&  & RAID & \textbf{0.1200} & \textbf{0.0530} & \textbf{0.1490} & \textbf{0.0600} & \textbf{0.1330} & \textbf{0.0600} & \textbf{0.1640} & \textbf{0.0680} & \textbf{0.1210} & \textbf{0.0540} & \textbf{0.1540} & \textbf{0.0610} \\
\midrule

\multirow{10}{*}{LFM-2B} & \multirow{5}{*}{Gender} 
& Original & 0.0270 & 0.0110 & 0.0330 & 0.0130 & 0.0370 & 0.0160 & 0.0470 & 0.0190 & 0.0440 & 0.0210 & 0.0520 & 0.0230 \\
&  & DP-PoT & 0.0070 & 0.0030 & 0.0080 & 0.0030 & 0.0030 & 0.0010 & 0.0050 & 0.0020 & 0.0080 & 0.0030 & 0.0090 & 0.0040 \\
&  & AU-PoT & 0.0120 & 0.0050 & 0.0160 & 0.0060 & 0.0280 & 0.0120 & 0.0330 & 0.0130 & 0.0400 & \textbf{0.0190} & 0.0490 & \textbf{0.0210} \\
&  & Adv-InT & 0.0220 & 0.0090 & 0.0270 & 0.0100 & 0.0290 & 0.0120 & 0.0360 & 0.0140 & 0.0390 & 0.0170 & 0.0440 & 0.0180 \\
&  & RAID & \textbf{0.0240} & \textbf{0.0100} & \textbf{0.0290} & \textbf{0.0110} & \textbf{0.0310} & \textbf{0.0140} & \textbf{0.0370} & \textbf{0.0160} & \textbf{0.0420} & 0.0180 & \textbf{0.0500} & 0.0200 \\ \cmidrule{2-15}
& \multirow{5}{*}{Age} 
& Original & 0.0270 & 0.0110 & 0.0330 & 0.0130 & 0.0370 & 0.0160 & 0.0470 & 0.0190 & 0.0440 & 0.0210 & 0.0520 & 0.0230 \\
&  & DP-PoT & 0.0060 & 0.0020 & 0.0080 & 0.0030 & 0.0030 & 0.0010 & 0.0050 & 0.0020 & 0.0070 & 0.0030 & 0.0090 & 0.0040 \\
&  & AU-PoT & 0.0210 & 0.0090 & \textbf{0.0280} & 0.0090 & 0.0280 & 0.0120 & 0.0340 & 0.0140 & \textbf{0.0410} & \textbf{0.0190} & \textbf{0.0490} & \textbf{0.0210} \\
&  & Adv-InT & 0.0220 & 0.0100 & 0.0270 & 0.0110 & 0.0290 & 0.0130 & 0.0370 & 0.0150 & 0.0390 & 0.0170 & 0.0450 & 0.0180 \\
&  & RAID & \textbf{0.0240} & \textbf{0.0110} & \textbf{0.0280} & \textbf{0.0120} & \textbf{0.0340} & \textbf{0.0160} & \textbf{0.0400} & \textbf{0.0180} & 0.0390 & 0.0170 & 0.0480 & 0.0200 \\
\midrule

\multirow{10}{*}{KuaiSAR} & \multirow{5}{*}{Gender} 
& Original & 0.0410 & 0.0180 & 0.0480 & 0.0200 & 0.0430 & 0.0180 & 0.0520 & 0.0200 & 0.0500 & 0.0210 & 0.0610 & 0.0240 \\
&  & DP-PoT & 0.0090 & 0.0040 & 0.0130 & 0.0050 & 0.0070 & 0.0030 & 0.0090 & 0.0030 & 0.0160 & 0.0070 & 0.0200 & 0.0080 \\
&  & AU-PoT & 0.0380 & 0.0160 & 0.0490 & 0.0190 & \textbf{0.0390} & \textbf{0.0180} & \textbf{0.0490} & \textbf{0.0200} & 0.0430 & 0.0190 & 0.0530 & 0.0210 \\
&  & Adv-InT & 0.0410 & 0.0180 & 0.0480 & 0.0200 & 0.0350 & 0.0160 & 0.0440 & 0.0180 & 0.0280 & 0.0120 & 0.0330 & 0.0140 \\
&  & RAID & \textbf{0.0430} & \textbf{0.0190} & \textbf{0.0550} & \textbf{0.0220} & \textbf{0.0390} & \textbf{0.0180} & \textbf{0.0490} & \textbf{0.0200} & \textbf{0.0470} & \textbf{0.0200} & \textbf{0.0580} & \textbf{0.0220} \\ \cmidrule{2-15}
& \multirow{5}{*}{Age} 
& Original & 0.0400 & 0.0170 & 0.0500 & 0.0200 & 0.0430 & 0.0180 & 0.0520 & 0.0200 & 0.0500 & 0.0210 & 0.0610 & 0.0240 \\
&  & DP-PoT & 0.0090 & 0.0030 & 0.0110 & 0.0040 & 0.0070 & 0.0030 & 0.0090 & 0.0030 & 0.0160 & 0.0060 & 0.0190 & 0.0070 \\
&  & AU-PoT & \textbf{0.0360} & \textbf{0.0150} & 0.0460 & \textbf{0.0180} & \textbf{0.0390} & \textbf{0.0170} & \textbf{0.0500} & \textbf{0.0200} & 0.0410 & 0.0170 & 0.0500 & 0.0200 \\
&  & Adv-InT & 0.0260 & 0.0100 & 0.0310 & 0.0110 & 0.0240 & 0.0090 & 0.0310 & 0.0110 & 0.0240 & 0.0100 & 0.0310 & 0.0120 \\
&  & RAID & \textbf{0.0360} & \textbf{0.0150} & \textbf{0.0470} & \textbf{0.0180} & 0.0380 & 0.0160 & \textbf{0.0500} & 0.0190 & \textbf{0.0440} & \textbf{0.0200} & \textbf{0.0560} & \textbf{0.0230} \\
\midrule

\multirow{10}[4]{*}{Last.fm-360K} & \multirow{5}[2]{*}{Gender} & Original & 0.1390  & 0.0620  & 0.1690  & 0.0690  & 0.1420  & 0.0640  & 0.1720  & 0.0710  & 0.1470  & 0.0660  & 0.1780  & 0.0730  \\
&  & DP-PoT & 0.1200  & 0.0520  & 0.1460  & 0.0590  & 0.1170  & 0.0510  & 0.1430  & 0.0570  & 0.0410  & 0.0180  & 0.0500  & 0.0200  \\
&  & AU-PoT & \textbf{0.1390 } & \textbf{0.0620 } & \textbf{0.1690 } & \textbf{0.0690 } & \textbf{0.1420 } & \textbf{0.0640 } & \textbf{0.1720 } & \textbf{0.0710 } & \textbf{0.1470 } & \textbf{0.0660 } & \textbf{0.1780 } & \textbf{0.0730 } \\
&  & Adv-InT & 0.1340  & 0.0600  & 0.1640  & 0.0670  & 0.1380  & 0.0610  & 0.1660  & 0.0680  & 0.1210  & 0.0550  & 0.1460  & 0.0610  \\
&  & RAID  & \textbf{0.1390 } & \textbf{0.0620 } & \textbf{0.1690 } & \textbf{0.0690 } & 0.1380  & 0.0620  & 0.1680  & 0.0690  & \textbf{0.1470 } & \textbf{0.0660 } & \textbf{0.1780 } & \textbf{0.0730 } \\
\cmidrule{2-15}          & \multirow{5}[2]{*}{Age} & Original & 0.1390  & 0.0620  & 0.1690  & 0.0690  & 0.1420  & 0.0640  & 0.1720  & 0.0710  & 0.1470  & 0.0660  & 0.1780  & 0.0730  \\
&  & DP-PoT & 0.1200  & 0.0520  & 0.1460  & 0.0590  & 0.1170  & 0.0510  & 0.1430  & 0.0570  & 0.0410  & 0.0180  & 0.0500  & 0.0200  \\
&  & AU-PoT & \textbf{0.1390 } & \textbf{0.0620 } & \textbf{0.1690 } & \textbf{0.0690 } & \textbf{0.1420 } & \textbf{0.0640 } & \textbf{0.1720 } & \textbf{0.0710 } & \textbf{0.1470 } & \textbf{0.0660 } & \textbf{0.1780 } & \textbf{0.0730 } \\
&  & Adv-InT & 0.1340  & 0.0590  & 0.1630  & 0.0660  & 0.1320  & 0.0570  & 0.1610  & 0.0640  & 0.1360  & 0.0620  & 0.1660  & 0.0690  \\
&  & RAID  & \textbf{0.1390 } & \textbf{0.0620 } & \textbf{0.1690 } & \textbf{0.0690 } & \textbf{0.1420 } & \textbf{0.0620 } & 0.1710  & 0.0690  & \textbf{0.1470 } & \textbf{0.0660 } & \textbf{0.1780 } & \textbf{0.0730 } \\
          
\bottomrule
\end{tabular}
}
\end{table*}


